\documentclass[a4paper, 12pt]{report}
\usepackage{reportSty}
\usepackage{graphicx}
\usepackage{amssymb}
\usepackage{titlesec}
\usepackage{setspace}
\usepackage{fancyhdr}
\usepackage{wrapfig}
\usepackage{array}
\usepackage{mathtools, nccmath}
\usepackage{algorithm}
\usepackage[noend]{algpseudocode}
\usepackage[most]{tcolorbox} 
\tcbuselibrary{theorems} 

\usepackage{hyperref}
\hypersetup{
    colorlinks=true,
    linkcolor=blue,       
    citecolor=blue,       
    filecolor=magenta,    
    urlcolor=blue         
}

\titleformat{\chapter}[display]
  {\normalfont\huge\bfseries}
  {\chaptertitlename\ \thechapter}
  {15pt} 
  {\Huge}
\titlespacing*{\chapter}{0pt}{-25pt}{50mm}

\titlespacing*{\chapter}{0pt}{-40pt}{20pt}  

\usepackage{caption}   
\usepackage{subcaption} 
\usepackage{amsmath}

\usepackage{float} 
\usepackage{epigraph} 
\usepackage{amsthm}
\newtheorem{definition}{Definition}
\usepackage[titletoc]{appendix}

\usepackage{glossaries}
\setglossarystyle{index}

\usepackage{listings}
\usepackage{color}

\definecolor{codegreen}{rgb}{0,0.6,0}
\definecolor{codegray}{rgb}{0.5,0.5,0.5}
\definecolor{codepurple}{rgb}{0.58,0,0.82}
\definecolor{backcolour}{rgb}{0.95,0.95,0.92}

\usepackage{algorithm}
\usepackage{algpseudocode}

\makeglossaries
\begin{document}

\newcommand{\condBreak}[1]{\ifx#1\empty\else \\ \fi }

\begin{titlepage}
    \begin{center}
    \setstretch{1.2}
        \vspace*{0.3in}
        
        \begin{figure}[H]
            \centering
            \includegraphics[width=0.43\textwidth]{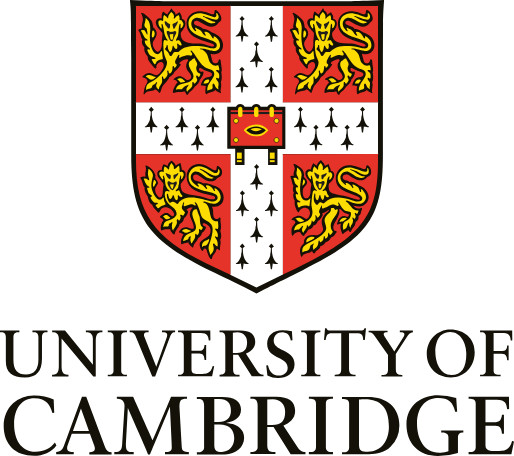}
        \end{figure}

        \vspace{0.5in}

        {\Huge \bf Musical Score Following using} \\
        \vspace*{.08in}
        {\Huge \bf Statistical Inference} \\
        \vspace*{.3in}
        {\Large{Josephine Cowley}} \\
        \vspace{0.2in}
        {\textbf{Supervisor} \\
                    Professor Simon J. Godsill            
                }
    
        \vspace{0.8in}

        {Submitted in partial fulfilment of the requirements
            \\for the Degree of MEng in Engineering}

        {on 27 May 2024 \\}

        \vspace{1.5in}

        {Cambridge University Engineering Department}

    \end{center}
\end{titlepage}

\titleformat{name=\chapter,numberless}
  {\centering\normalfont\Huge\bfseries} 
  {}
  {0pt}
  {}
\titlespacing*{name=\chapter,numberless}
  {0pt}{15pt}{40pt} 

\chapter*{\centering {Musical Score Following using Statistical Inference}}
\addcontentsline{toc}{chapter}{Technical Abstract}
\vspace{-30pt} 
{
  \centering
  \vspace{0.17in}
  \textbf{Josephine Cowley} \par
  St John's College, Cambridge \par
  Supervisor: Professor Simon J. Godsill\par  
}

\vspace{0.15in}

\subsection*{\centering \uppercase{Technical Abstract}}
Musical score following is the real-time mapping of a performance to corresponding locations in a musical score. Score following can be used in a variety of applications including automatic page turning and real-time accompaniment. 
This report presents a novel approach for score following motivated by Wilson and Adams's 2013 paper, which introduces Spectral Mixture (SM) kernels for Gaussian Process (GP) regression. Since the SM kernel is derived from a Mixture of Gaussians in the frequency domain, it is particularly suitable for modelling the superposed power spectra of musical notes, in which energy is concentrated at multiples of the fundamental frequency of each note.   
Our score follower begins by using a GP to statistically infer the musical notes played during 800-sample `audioframes' ($\approx$18 ms) of solo piano music. 
These predictions are then used in a duration-dependent Hidden Markov Model to predict the most likely score positions in real time. Our two-stage approach achieves successful score following not only on four-part hymns arranged for keyboard, but also on pieces for the violin, oboe, and flute. This showcases the powerful and flexible nature of GPs for statistical inference on musical audio signals. Given the success of this project, we contribute to the literature a first proof of concept of the application of GPs in score
following, and more broadly, in online Music Information Retrieval (MIR) tasks. This project also contributes a working score follower product that renders score position in real time using an adapted open-source user interface. Areas for future work include improving accuracy on repeated notes and during heavy use of sustain pedal, adapting to minor deviations from the score, and modelling multi-instrument works.


\subsection*{\centering Report Structure}

The report has five parts: \textit{Introduction} (chapters 1–3); \textit{Background} (chapters 4–5);  \textit{Statistical Inference} (chapters 6–8);  \textit{Implementation} (chapters 9–10); and  \textit{Discussion} (chapter 11). \\

\textbf{Part I, \textit{Introduction}}, introduces central concepts for our project. In chapter 1, we open with a brief description of score following and our motivations, as well as a statement of the project goal. In chapter 2, we present important musical preliminaries, exploring the physical phenomena and underlying score features that affect how music is perceived. We also introduce some common standards for music notation. In chapter 3, we formally define the problem of score following. We then take inspiration from manual score following, contrasting it with the challenges faced by automatic score followers. We close with a general framework for score followers, which sets us up for the literature review in Part II and designing our own score follower in Part III.   \\

\textbf{Part II, \textit{Background}}, sets the scene for score following. In chapter 4, we detail several key applications for score following alongside some commercial products. We then carry out a literature review. First, we investigate general score following by chronologically tracing the development of score following over the last four decades. We then examine existing methods for pitch detection and note their drawbacks to motivate our use of GPs. We then perform a literature review of GPs in the context of MIR. Finally, we outline the key recurring challenges from this research. In chapter 5, we lay out the mathematical notation and framework for GPs and detail the Spectral Mixture kernel, which we use in our GP model. \\

\textbf{Part III, \textit{Statistical Inference}}, details the statistical and algorithmic methods we use for score following. In chapter 6, we present a general methodological overview and introduce the high-level framework of our score follower, which contains two critical stages: \textit{GP Model Specification} and \textit{Real-Time Alignment}.  In chapter 7, we commence with Stage 1: \textit{GP Model Specification}. The tasks of this chapter include analysing and modelling musical signals, designing a covariance matrix, tuning our GP's hyperparameters and formulating a log marginal likelihood (LML) function. Finally, we present the results from this chapter, which were largely successful, showing accurate underlying frequency prediction. Chapter 8 focuses on Stage 2: \textit{Real-Time Alignment} and details the statistical inference methods for score following. We use a Hidden Markov Model (HMM) to represent audioframes as observed variables that depend upon latent states (i.e., the underlying notes from the score). We use the LML function from Chapter 7 for the HMM's emission probabilities, and develop a state duration model that uses rhythmic information from the score for transition probabilities. We then implement an online `Windowed' Viterbi algorithm to achieve efficient and near-optimal score following by computing the most likely sequence of latent states. To close, we discuss the results, which were largely positive except for tests where there was excessive use of sustain pedal. \\

\textbf{Part IV, \textit{Implementation}}, presents the final product. In chapter 9, we first outline the framework of the implemented score follower and its modes of operation. We then describe the final product in terms of its components, detailing constituent system architectures and design justifications. In chapter 10, we present the key results of our  score follower, evaluating its performance on several different test cases. Our score follower was able to follow an intermediate monophonic piano piece, a simple 2-part piece and even a 4-part hymn. The follower also performed excellently on various instruments, including the violin, flute and oboe, demonstrating the flexibility of our GP model. The main limitations identified were use of sustain pedal, deviations from the score and repeated notes or passages. \\

\textbf{Part V, \textit{Discussion}}, consists of only chapter 11, which summarises our project. We detail key findings from Parts III and IV, covering the main contributions of this report. We wrap up by assessing areas for future work and conclude with some final reflections.


\titleformat{name=\chapter,numberless}
  {\normalfont\Huge\bfseries} 
  {}
  {0pt}
  {}
\titlespacing*{name=\chapter,numberless}
  {0pt}{-40pt}{25pt} 

\vbox{
    \begin{minipage}[t][1\textheight][t]{\textwidth}
        \chapter*{\huge Acknowledgements}

        \addcontentsline{toc}{chapter}{Acknowledgments}

        First, I would like to thank Professor Simon J. Godsill for his support and guidance throughout this project. I have appreciated our invaluable discussions at regular supervisions, where often he would set me back on track whenever obstacles were encountered. Second, a grateful acknowledgment to Professor Carl Rasmussen for one supervision where he offered his expertise on Gaussian Processes.\\

        It goes without saying that I am indebted to my mother and father, who have provided the love and support that has got me here. Of course, a special thanks to my two sisters for being my best friends. \\

        The course of this year has certainly been challenging at times, and I'd like to say thanks to all my dear friends who have encouraged me all the way. Lastly, a big thank you to Patrick, whose unending care and support has especially helped throughout the course of this project.

    \end{minipage}



}

\newglossaryentry{articulation}{
   name=articulation,
   description={defines how smoothly notes are played. One type of articulation marking is \gls{legato}}
}

\newglossaryentry{homophonic}{
   name=homophonic,
   description={music refers to a type of musical \gls{texture} which has one line of melody being played by multiple lines simultaneously}
}

\newglossaryentry{key}{
   name=musical key,
   description={refers to the main notes, scales and chords which a section of music is built from}
}

\newglossaryentry{legato}{
   name=legato,
   description={is a playing technique where a musician smoothly connects notes. The opposite is \textit{staccato} where notes played are detached}
}

\newglossaryentry{monophonic}{
   name=monophonic,
   description={music refers to a type of musical \gls{texture} which consists of a single line or melody}
}

\newglossaryentry{ornament}{
   name=ornaments,
   description={are added notes that are not essential to carry the overall line of the melody (or harmony), but serve instead to decorate a musical line}
}

\newglossaryentry{phrasing}{
   name=phrasing,
   description={is the method in which a musician shapes passages of music to allow for expression}
}

\newglossaryentry{polyphonic}{
   name=polyphony,
   description={refers to a type of musical \gls{texture} which consists of two or more simultaneous lines of independent melody}
}

\newglossaryentry{rubato}{
   name=rubato,
   description={refers to \gls{tempo} deviations musicians may add for expression}
}


\newglossaryentry{semiquaver}{
   name=semiquavers,
   text=semiquaver,
   description={are notes played for $\frac{1}{16}$ of the duration of a semi-breve (the second-longest note value used in modern staff notation). This means they tend to have rather short durations (though this also depends on the underlying \gls{tempo})}
}
\newglossaryentry{demisemiquavers}{
   name=demisemiquavers,
   description={are notes played for $\frac{1}{32}$ of the duration of a semi-breve (the second-longest note value used in modern staff notation). This means they tend to have  very short durations (though this also depends on the underlying \gls{tempo})}
}

\newglossaryentry{sustain pedal}{
   name=sustain pedal,
   description={is a keyboard technique for allowing the strings of the instrument to openly resonate, rather than being damped}
}

\newglossaryentry{texture}{
   name=texture,
   description={refers to the structure of different layers of sound in a piece, and the relationship between them}
}

\newglossaryentry{timbre}{
   name=timbre,
   description={characterises the different acoustic sound qualities that distinct instruments exhibit}
}

\newglossaryentry{glissandi}{
   name=glissando,
   description={is a technique where the instrumentalist slides between notes}
}

\newglossaryentry{tempo}{
   name=tempo,
   description={is the speed of a performance. A tempo marking gives the composer's intended tempo, measured in beats per minute}
}


\newglossaryentry{dynamics}{
   name=dynamics,
   description={define the subjective loudness at which the composer intended marked music passages to be played}
}

\newglossaryentry{harmonic progression}{
   name=harmonic progression,
   description={(or chord progression) is the movement from one chord to another which builds a structural foundation for harmony and \gls{key}}
}

\newglossaryentry{bars}{
   name=musical bars,
   text=bar,
   description={are segments of music bounded by vertical lines known as bar lines. Bars contain a constant number of beats, determined by the piece's \textit{time signature}}
}

\newglossaryentry{rhythm}{
   name=rhythm,
   description={is the arrangement of sounds and silences in time in a perceptibly structured way, often containing a pulse or beat}
} 

\printglossary[title=Glossary of Musical Terms]{\label{glossary}}
\addcontentsline{toc}{chapter}{Glossary of Musical Terms}

\tableofcontents


\newpage
\setstretch{0.95}
\setcounter{page}{1}
\pagenumbering{arabic}

\part{I. Introduction}
\chapter{Introduction}{\label{ch:intro}}
With the proliferation of portable digital devices such as iPads and tablet computers, musicians have begun to utilise these technologies as tools for music-making. Most notably, digital scores offer the practical advantage of storing a musician's entire repertoire, replacing dozens of books and scores. Free online catalogues such as \textit{IMSLP}\footnote{\href{https://imslp.org/}{https://imslp.org/}} provide free instant access to scores, and applications such as \textit{forScore}\footnote{\href{https://forscore.co/}{https://forscore.co/}} provide an integrated library, organisation and practice tool. Hence, these devices are rapidly becoming the norm in practice rooms and are even making their way into concert halls.\\

\begin{figure}[H]
    \centering
    \includegraphics[width=0.7\textwidth]{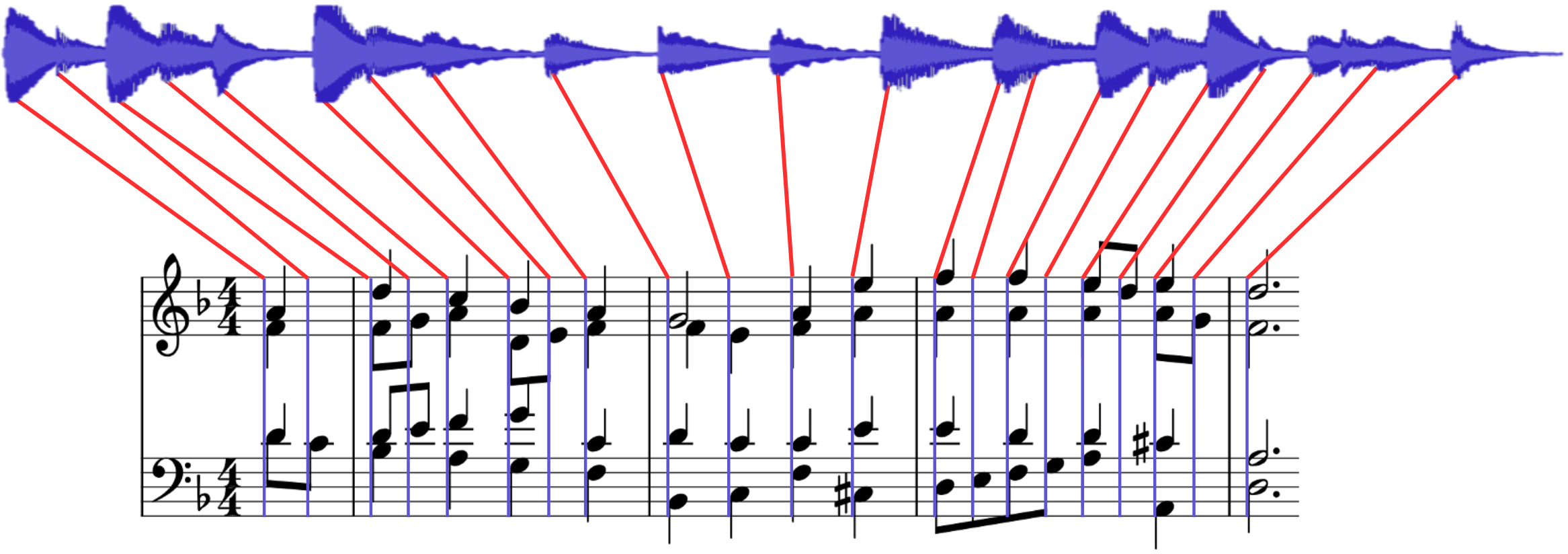}
    \caption{Illustration of score following: positions (blue) in the musical score are mapped to locations in a live-streamed recording (red) of \textit{O Haupt voll Blut und Wunden} arranged by Bach.}
\label{fig:score_follower}
\end{figure}

A logical next step would be for our digital devices to go beyond merely replacing sheet music. These devices have the potential to become an integral part of interactive practice, analysis and feedback. Often, the task of \textit{score following} is a first step in the real-time analysis of a musical performance. This involves mapping real-time positions of a performance to corresponding locations in a score, as depicted in \hyperref[fig:score_follower]{Figure \ref*{fig:score_follower}}. Despite a plethora of research dating back as far as the 1980s, the lack of adoption of score followers reflects the limited performance of these products. This is because score following is no trivial task, and any successful application would need to be highly accurate.

\section{Motivation}
It is critical to use a statistical approach for score following due to deviations from the score, which are caused both by inevitable performance errors and deliberate interpretative decisions. Although there exist many statistical approaches for score following (see \hyperref[ch:score_following_literature_review]{chapter \ref*{ch:score_following_literature_review}}), this work is the first attempt in the literature that uses Gaussian Processes (GPs) for this application. A crucial component of GPs is their kernel, which can be fine-tuned to encode the properties of the signals to be modelled. Accordingly, GPs offer a flexible and robust solution for incorporating prior knowledge about the highly structured properties of music. Furthermore, GPs have been successfully used in music transcription and source separation, suggesting they may prove fruitful in the task of score following \cite{miscdo_2019_sparse}. Motivated by Wilson and Adams’s paper \cite{wilson_2013_gaussian}, which introduces Spectral Mixture Kernels for GP Regression, we use GPs to infer the likelihoods that certain notes were played during an audioframe (a group of contiguous audio samples). These predictions can then be used in a second statistical model that predicts score location over time from the note likelihoods. Thus, we develop a two-stage score follower which uses a GP in its first half.

\section{Project Goal}{\label{section:project_goal}}
The overall goal of this project is to use GPs to perform score following
on simple solo piano pieces. This is not only intended as a proof of concept that GPs can be used for real-time audio processing, but it is also a first step towards a usable score following product, demonstrating how GPs can be incorporated into real user-facing applications. Because of the lack of objective criteria for the success of score followers, we will use the subjective judgement of expert musicians as our primary benchmark. This is especially justified considering that score followers are ultimately to be used by musicians, and should meet their expectations. 

\section{Note to the Reader}
This report assumes a basic understanding of music theory, which includes the musical terms defined in the \hyperref[glossary]{glossary\ref*{glossary}}. These terms are highlighted in blue and linked to their definitions in the glossary upon first occurrence per section. We also assume a basic understanding of Bayesian statistics and GPs. 
\chapter{Music Preliminaries}{\label{ch:music_preliminaries}}

An investigation of score following requires an understanding of the physical properties of music and standard notation practices. This chapter first outlines the perceptual properties of sound and discusses the physical bases for the distinction between musical and non-musical sounds. Then, we asses which of these features are present in the score and are therefore relevant to score following. Finally, we introduce three different standards of musical notation and settle on using symbolic formats like MIDI and MusicXML in this application.

\section{Perception of Music}{\label{subsection:perception_of_music}}
When a trained musician listens to a performance, they make sense of the music by characterising pitch, duration, volume and \gls{timbre} \cite{donnelly_2015_learning}. These features can be explained in terms of underlying physical properties, as well as features from the score and player interpretation. 

\subsection{Perceptual Properties and Physical Correlates}{\label{section:duration}}

\textbf{Pitch} is the perceptual quality of music that makes notes sound higher or lower. Although pitch is determined by \textit{frequency spectra} (a physical phenomenon), different frequency spectra may be perceived as the same pitch. When a pitched instrument plays a note, we perceive the spectrum of frequencies as a single pitch. This frequency spectrum contains energy at the fundamental frequency $f_0$ and at \textit{overtones} (or \textit{partials}). \\ 

\begin{figure}[H]
    \centering
    \includegraphics[width=0.75\textwidth]{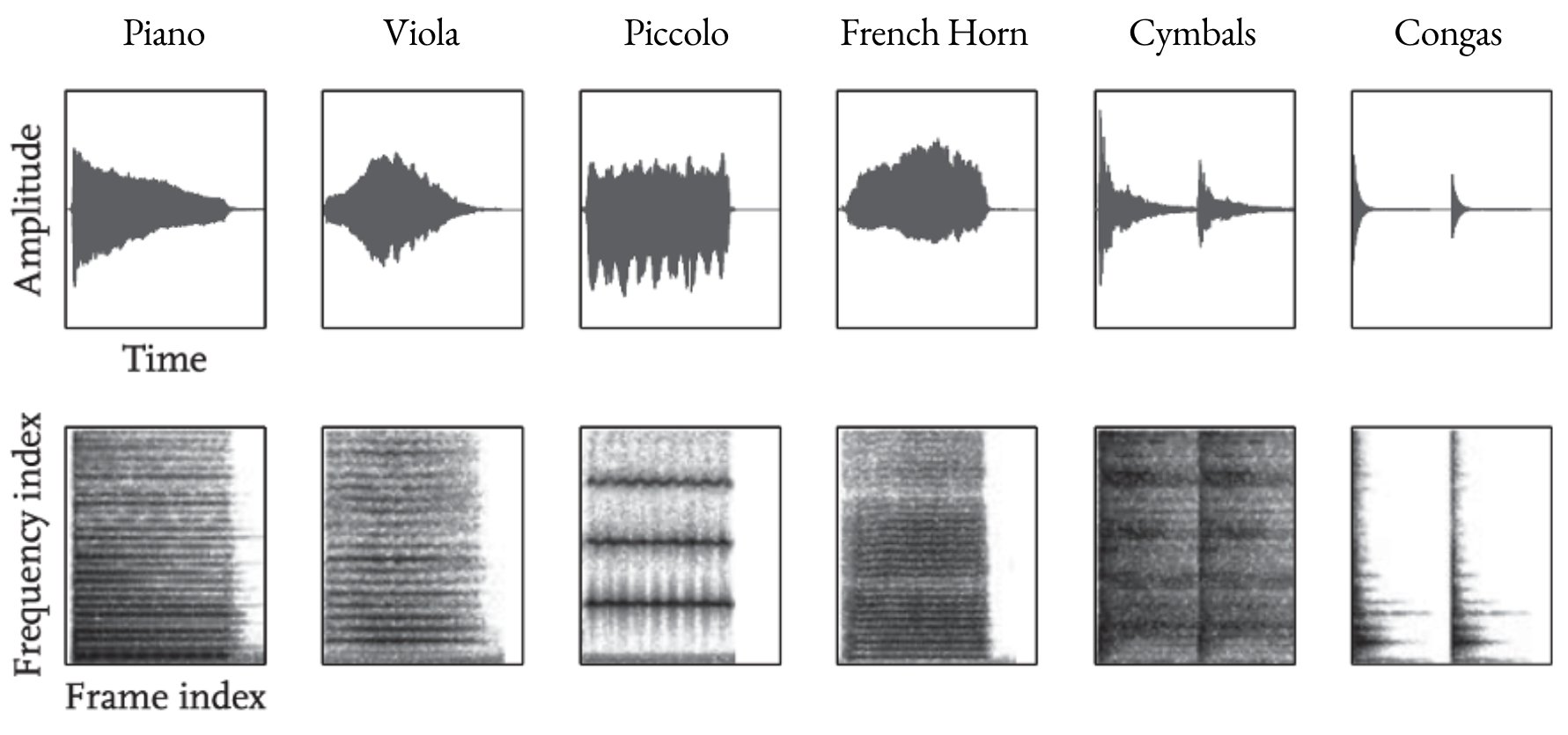}
    \caption{Examples from \cite{godsill_2006_bayesian} of different instruments' temporal envelopes (top) and spectrograms (bottom). Spectrograms represent the time-varying magnitude spectra—i.e. the modulus of the short-time Fourier transform. Audio data and images are from the RWCP Instrument samples database.}
    \label{fig:musical_properties}
\end{figure}

\textbf{Duration} is the length of a note’s amplitude-time waveform shape, which is also called the \textit{temporal envelope}. Percussive instruments like conga drums exhibit immediate amplitude decay, whereas sustaining instruments have slowly varying amplitude. The first row of graphs in \hyperref[fig:musical_properties]{Figure \ref*{fig:musical_properties}} shows different instruments' temporal envelopes. \\

Due to psychoacoustics, the perception of \textbf{loudness} is not only dependent on energy or amplitude, but also on frequency and waveform shape \cite{donnelly_2015_learning}. \\

A musical note's \textbf{timbre}, or `tone colour', is associated with how its sound is produced. For instance, a flute, bowed violin, and plucked violin all sound different. One underlying property is the shape of an instrument's \textit{power spectrum}, or its \textit{spectral envelope}. Unlike non-pitched instruments like untuned percussion, pitched instruments exhibit a \textit{harmonic series}, which concentrates energy at multiples of $f_0$. Even then, pitched instruments sound different from one another because the distribution of energy at those harmonics varies greatly. For instance, some instruments exhibit more pronounced even-numbered harmonics. The variation of spectral envelopes over time can be represented by \textit{spectrograms}, like those in the second row of \hyperref[fig:musical_properties]{Figure \ref*{fig:musical_properties}}. Additionally, a note's \textit{temporal envelope} influences its timbre.  \\

All four of these perceptual properties are also affected by the acoustic environment, since resonance, position and noise strongly influence our auditory perception. \\

\subsection{Influences from the Score}{\label{subsection:score_influences}}
We now turn to how the score influences the above perceptual features in music. We also consider the effects of unexpected deviations from the score, highlighting the limitations of using certain features for score following. \\

A score dictates which \textbf{pitches} are intended to be present by specifying musical notes at times. However, the pitches present in a performance may deviate due to performer mistakes. Additionally, performers may embellish a piece with \gls{ornament} or even improvisation. Finally, playing techniques such as \gls{sustain pedal} and \gls{legato} causes pitches from previous notes to continue to sound even after their end in notation. \\

Similar to pitch, note \textbf{durations} are strongly influenced by the score, which specifies \gls{rhythm}, \gls{tempo} and \gls{articulation}. Although some scores have exact tempo markings, many only have qualitative indications (e.g., \textit{Andante} means `walking pace'). Additionally, musicians are often expected to add \gls{rubato} for expression, which distorts rhythm in unspecified ways.\\

Composers specify \gls{dynamics} in the score to describe the intended \textbf{loudness} of certain notes. However, there is no objectively correct loudness given some dynamic. Rather, the loudness of a note is determined by the performer's \textit{interpretation} of those dynamics. Furthermore, musicians alter loudness through \gls{phrasing} or articulation for expression. Thus, loudness differs from pitch and duration in that there is no quantitative representation in the score corresponding to the perceptual feature in question.  \\


Similar to loudness, composers specify \textbf{\gls{timbre}} with performance directions, which are even more of a matter of subjective interpretation. One exception is that in a multi-instrument work, the choice of instrument predictably influences the resulting timbre. However, since we focus on the solo piano, we cannot make use of this information.\\ 

Overall, since only pitch and duration correlate strongly with the score, they are the only features we can reliably use to determine score position. This constrains our methods for developing a score follower, and also guides our choice of score notation formats.

\section{Music Notation}
\subsection{Modern Staff Notation}
There are many forms of musical notation, but \textit{modern staff notation} (or \textit{Western music notation}) is commonly used in different genres around the world. We therefore use this notation throughout this report, as seen in \hyperref[fig:score_follower]{Figure \ref*{fig:score_follower}}. 
Even in the digital age, sheet music is stored in visual formats via PDFs and images. However, most score followers do not work directly with these formats,\footnote{Only recently, due to advances in deep learning, have there been attempts to use image processing for alignment directly from sheet music \cite{10.3389/fcomp.2021.718340}.} but rather \textit{symbolic} representations of sheet music like MIDI or MusicXML. 

\subsection{MIDI Notation}{\label{subsection:midi}}
MIDI (\textit{Musical Instrument Digital Interface}\footnote{ \href{https://midi.org/midi-2-0}{https://midi.org/midi-2-0}}) is a technical standard for electronic instruments to communicate. It is also used to efficiently represent both audio recordings and certain score data. MIDI reduces score information down to a few salient features which correspond to the four perceptual features except \gls{timbre}: MIDI note numbers (pitch), note velocity (loudness), and note onsets and offsets (duration), meaning file sizes can be very small. Other score information, like layout and performance indications, is lost.

\subsection{MusicXML}
\textit{MusicXML}\footnote{\href{https://www.musicxml.com/}{https://www.musicxml.com/}} is a feature-wise superset of MIDI. However, MusicXML can also contain information such as the layout of a score in modern staff notation and performance indications. 

\subsection{Choice of Notation}{\label{subsection:choice_notation}}
Despite the expressive benefits of the MusicXML format over MIDI, we choose to use MIDI as the input for our score follower, since it is still the most universally available symbolic format for scores. Moreover, as we have established above, there are only two properties of the score we need for score following—namely pitch and duration—both of which are fully specified in MIDI. However, displaying a fully detailed score requires more information about the score layout and performance directions, and this information is only available in MusicXML. Therefore, the score renderer (see \hyperref[section:renderer]{section \ref*{section:renderer}}) uses MusicXML, which allows conversion to a fully detailed score in modern staff notation. Thus, after connecting the score follower to the score render, we can easily allow trained musicians to evaluate the score follower in modern staff notation.

\chapter{Problem Definition}{\label{ch:problem_definition}}

This chapter defines the problem of score following. We begin by briefly discussing why score following is a challenging problem, especially when contrasted with score alignment. Next, we examine the techniques used in manual score following (done by human musicians) and assess which are viable in automatic approaches to score following. Finally, we outline a general three-step approach to score following, which is used in the next chapter to discuss the existing literature, as well as to design our implementation.

\section{Defining Score Following}
We will adopt the general definition proposed by Arshia Cont, a prominent figure affiliated with IRCAM,\footnote{\href{https://www.ircam.fr/}{https://www.ircam.fr/}} who offers a concise definition of score following \cite{cont_arshia}:

\begin{definition}
   Score following serves as a real-time mapping interface from Audio abstractions towards Music symbols and from performer(s) live performance to the score in question.
\end{definition}

\subsection{Difficulties for Score Following}

Score following is a non-trivial task because every live performance of the same piece, even by the same performer, may vary significantly. As detailed in \hyperref[subsection:perception_of_music]{subsection \ref*{subsection:perception_of_music}}, the score at best partially determines what occurs in a performance of a work, not to mention performer error and acoustic phenomena external to the performance. The space of all possible events that can be considered performances of, say, Beethoven's \textit{Moonlight Sonata} is incredibly diverse, yet a robust score follower must be able abstract from irrelevant differences and focus on the common features derived from the score.


\subsection{Score Following Vs Score Alignment}{\label{subsection:score_following_v_alignment}}
Score following is not to be confused with score alignment. Score alignment assumes access to the entire performance recording, relaxing the real-time constraint of score following. Therefore, score alignment poses a lesser challenge, since computation time is not critical, and inference can be performed over all the available data. By contrast, for score following, efficiency is highly important, and estimations of score position must be constantly updated upon receiving new data.\\

The greater flexibility of score alignment means it has been heavily researched, yielding many robust solutions. Although score following has also long been researched, and several commercial products have been developed (see \hyperref[ch:score_following_literature_review]{chapter \ref*{ch:score_following_literature_review}}), none have proved reliable enough to receive broad uptake among musicians.

\section{Manual and Automatic Score Following}{\label{section:score_following_approaches}
\subsection{Manual Score Following}{\label{subsection:manual_score_following}}
During manual score following, human musicians can effectively utilise all four perceptual features of music: not just pitch and duration, but also loudness and timbre. This is because, in addition to the score, human musicians also have knowledge of the conventions of music performance, subjective memories of loudness and timbre, emotional responses corresponding to performance directions, and in some cases even visual cues. Despite—or because of—this wealth of information, musicians do not consider all these features simultaneously. Rather, they focus their attention on prominent score features that they intuit will help locate themselves within the piece. For instance, rather than trying to track every pitch present, a musician may focus on overall melodic contour (i.e., rising and falling patterns in note pitch), certain noticeable lines (e.g. melody or bass lines), or prominent \gls{harmonic progression}s. Similarly, musicians are likely to use an innate sense of rhythm to count beats of \gls{bars}s to help identify \gls{phrasing}, abstracting from the rhythmic detail of single notes. All this requires knowledge and understanding of music and music theory.


\subsection{Automatic Score Following}
Automatic score followers face three distinct challenges compared to manual score followers. First, we have established that automatic score followers are largely limited to using only pitch and duration (see \hyperref[subsection:score_influences]{subsection \ref*{subsection:score_influences}}), since the perceptual features of loudness and \gls{timbre} do not have quantitative representations in the score and are therefore more difficult for a computer to use. Second, while skilled manual score followers use musical understanding to pay attention to only contextually relevant features, this high-level understanding is impossible to imperatively encode.\footnote{Currently, only deep learning techniques offer methods for learning these intuitions. E.g., see \cite{10.3389/fcomp.2021.718340}.} Finally, while human score followers directly perceive pitch and rhythm, computers do not, and must be programmed to infer these features from raw audio. \\

To counter some of the variation caused by irrelevant underlying physical properties (e.g. \gls{timbre}), as well as the possibility of background noise, we design our model for solo piano and assume the audio data contains no significant sounds other than the piano. 

\subsection{General Approach}{\label{subsection:general_approach}}
 Given the lesser resources available to automatic score followers, score followers typically need to follow a three-step approach: \textbf{feature extraction, similarity calculation}, and \textbf{alignment} \cite{chen_2018_an}. Feature extraction involves inferring pitch and rhythmic components from audio. This step mimics basic human auditory perception. Next, similarity calculation looks for correspondences between these features and particular locations in the score. This step requires the ability to read the score's notes and match them to pitches and rhythms. Finally, alignment makes a single best estimate of score position from a sequence of these correspondences. This requires a functional understanding of how single locations in a score `fit together' in context. Different techniques used at each step depends on various implementations, as presented in the next chapter.

\part{II. Background}
\chapter{A Review of Score Following}{\label{ch:score_following_literature_review}}

In this chapter we set the scene for score following. We commence by presenting a few applications, primarily detailing automatic page turning (APT) and computer-aided accompaniment (CAA). We then carry out a literature review, which begins with historical approaches to general score following, followed by an overview of general pitch detection techniques. The literature review then turns to Gaussian Processes (GPs) in Music Information Retrieval (MIR). Finally, we consider challenges that have been identified in the literature.   

\section{Applications and Commercial Products}
Score following is used in a plethora of potential applications and commercial products:

\subsection{Automatic Page Turning} {\label{subsection:APT}}
Manual page turning during a live performance is a highly specialised skill that demands sight-reading aptitude, constant concentration and an understanding of pre-agreed performance decisions, such as repeats. Therefore, it is not uncommon for concerts to suffer slip-ups due to page turner mistakes.  However, performances make up only a small fraction of the time that page turning is needed. Most page turning occurs in the practice room, where hiring a page turner is infeasible. Thus, page turning is a common source of frustration. \\

The demand for APT solutions is demonstrated by the myriad of commercial products available. However, none use \textit{true} score following. \textbf{Manual APT Devices} rely on gestures from the player to turn the page. Many applications use a physical device, like a foot pedal, to transmit BLE messages to the primary device which turns the page. Examples include \textit{Stomp Bluetooth 4.0}\footnote{\href{https://www.codamusictech.com/products/bluetooth-page-turner-music-pedal-for-tablets}{https://www.codamusictech.com/products/bluetooth-page-turner-music-pedal-for-tablets}} and \textit{Airturn}.\footnote{\href{https://www.airturn.com/}{https://www.airturn.com/}} Though simple and intuitive, manual APT devices distract performers and introduce a new potential point of performance interruption.  \textbf{Scrolling APT Applications} do not require the musician to initiate turns, since they employ a score-scrolling feature. The scroll rate is determined by the \gls{tempo} of a pre-recorded playback of the piece, and therefore scrolling APT applications cannot accommodate spontaneous changes in tempo or unexpected score deviations. Examples include \textit{MobileSheets}\footnote{\href{https://www.zubersoft.com/mobilesheets/}{https://www.zubersoft.com/mobilesheets/}} and \textit{Musicnotes}.\footnote{\href{https://www.musicnotes.com/apps/}{https://www.musicnotes.com/apps/}}


\subsection{Computer-Aided Accompaniment}
Individual musicians often depend on others—from a single keyboard accompanist to a small ensemble to a whole symphony orchestra—to make music. Computer-aided accompaniment (CAA) aims to provide the supporting lines of a piece such that a single musician can play a multi-instrument work.  Several commercial CAA products exist, including Yamaha’s\textit{CueTIME}\footnote{\href{https://hub.yamaha.com/pianos/p-digital/cuetime-the-software-that-follows-you/}{https://hub.yamaha.com/pianos/p-digital/cuetime-the-software-that-follows-you/}} which provides pre-recorded MIDI instrumentals that ‘wait’ for performer keying information. However, CueTIME is limited to certain Yamaha keyboards. 


\subsection{Other Applications}
Other score following applications include: performance cues for lighting and camera equipment in live shows; music teaching and pedagogical reasons, such as score study; guided practice, for indication of mistakes; and entertainment, such as karaoke. 




\section{Literature Review}{\label{section:literature_review}}
\subsection{General Approaches to Score Following: Three Eras}

The history of score following can be divided into three eras \cite{lee_2022_final}. The first era largely addressed computer aided accompaniment. In 1984, Dannenberg \cite{dannenberg_1984_algorithm} and Vercoe \cite{vercoe_1984_the} independently presented string-matching approaches using pitch detection. Dannenberg’s group later developed this work to make it robust to \gls{polyphonic}, performance error, \gls{rubato} and \gls{ornament} \cite{dannenberg_1988_new}\cite{bloch_1985_realtime}. Baird went on to develop this into phrase-matching, where whole musical phrases were matched rather than individual notes \cite{baird_1990_the}\cite{baird_1993_artificial}. In 1995, Vantomme introduced the use of temporal features for situations where pitch-based methods were inadequate \cite{vantomme_1995_score}. \\

The second era of score following accommodated performance error using stochastic models. Cano et al.’s work was the first score follower to use a Hidden Markov Model (HMM) and used energy, zero-crossings and fundamental frequency as the observed emissions \cite{cano_1999_scoreperformance}. In 1999, Raphael developed a HMM which instead used spectral features \cite{raphael_1999_automatic}. This work proved seminal, forming the foundation of much research and a commercial product \cite{raphael_2006_aligning}. Dynamic Time Warping (DTW) approaches were also developed during this time in \cite{orio_2001_alignment} and \cite{dixon_2005_match}. \\

Finally, a number of different approaches have emerged since the 2010s. For instance, Shinji et al.’s group explored the use of Conditional Random Fields, and several papers have used Sequential Monte-Carlo methods (Particle Filtering) \cite{sako_2014_ryry}\cite{yamamoto_2013_robust}. Some uniquely interesting approaches include a multimodal neural network that takes images of sheet music \cite{matthiasdorfer_2016_towards} and an eye-tracking technique presented in \cite{noto_2019_adaptive}.

\subsection{General Approaches to Pitch Detection: Three Categories}
The numerous approaches to pitch detection fall into three categories: time domain, frequency domain, and hybrid approaches \cite{McLeod2008FastAP}. One of the simplest time domain methods is zero-crossing, such as in Cooper and Ng's \cite{zero_crossing}. In 1969, Gold and Rabiner developed a parallel processing technique in \cite{Gold_Rabiner}, which was further developed as a real-time algorithm by Sukkar et al. in \cite{Sukkar}. Autocorrelation methods have been used extensively in both music and speech processing, such as in \cite{Rabiner}. In 1993, Jacovitti and Scarano used an Average Squared Difference Function \cite{Jacovitti}, which de Cheveign and Kawahara further developed in 2002 \cite{Yin}. \\

In the frequency domain, there are many spectral peak estimators, including the peak detection algorithm in McLeod and Wyvill's \cite{Mcleod} and Keiler and Marchand's parabolic fitting technique \cite{Keiler}. In 1999, Wakefield used chromagrams for vocal pitch detection \cite{wakefield99_maveba}. Constant-Q transform coefficients, introduced in \cite{Youngberg}, have been used in several more recent papers \cite{Nakamura}\cite{Chen}. Mel Frequency Ceptral Coefficients were explored in \cite{Logan2000MelFC} for music, rather than in the field of speech processing where it is widely used. \\

Finally, there are several hybrid and other pitch detection techniques, including wavelet transforms \cite{7d015f6ce44c4b199dae7db41389965a}, Cepstrum  analysis \cite{Noll}, and Non-negative Matrix Factorisation \cite{godsill_2006_bayesian}.\\

A common drawback, especially among time domain approaches, is that they do not provide statistical likelihoods for detected pitches. This makes their results difficult to use in higher-level statistical inference models, like those used in score following. This motivates the investigation of Gaussian Processes (GPs) as a time domain method that yields likelihoods of pitches, which can be naturally incorporated in a Bayesian framework for score following.       

\subsection{Gaussian Processes for Music Information Retrieval}
Although there have been several explorations of using GPs to model audio signals \cite{alvarado_2016_gaussian}\cite{wilkinson_2019_gaussian}, this is a novel exploration of using GPs for score following. In 2014, Turner and Sahani used GPs for inferring superposed source components \cite{turner_2014_timefrequency}. Similarly, Liutkus et al. performed undetermined source separation involving GP regression \cite{liutkus_2011_gaussian}. GPs have also been used for jointly estimating the spectral envelope and fundamental frequency of a speech signal, as well as time-domain audio source separation \cite{yoshii_2015_masataka}\cite{yoshii_2013_beyond}. Alvarado and Stowell are particularly relevant contributors to this field. In 2016, they used GPs for pitch estimation and inferring missing segments in a polyphonic recording \cite{alvarado_2016_gaussian}. Their later work presented sparse GPs for source separation using spectrum priors in the time-domain \cite{miscdo_2019_sparse}.

 \section{Challenges}{\label{section:challenges}}
A number of challenges recur throughout the literature:
\begin{itemize}
    \item Player deviation from the score may occur due to mistakes or improvisation. 

    \item Background noise and acoustical effects such as resonating strings can cause issues for pitch detection. 
    \item Score following must work in real-time, and therefore must not overload the CPU. 
    \item Thick \gls{texture}s (e.g. \gls{polyphonic}) increase the complexity of the problem, since there are more simultaneous pitches to detect. 
    \item Evaluation of score followers is not well-standardised. Hence, it is difficult to compare the performance of a score follower to others in the literature.   
\end{itemize}

\chapter{Gaussian Processes}{\label{ch:Gaussian Processes}}
This chapter introduces Gaussian Processes (GPs) and shows how careful choice of covariance function (or kernel) allows us to encode assumptions about distributions over functions. In particular, we detail the Spectral Mixture (SM) covariance function, which allows modelling signals as superposed periodic components. All of the equations and definitions in this chapter come from \cite{rasmussen_2006_gaussian}. 

\section{Definition}
GPs provide a principled, practical and powerful framework for non-parametric regression and classification. They can be thought of as generalisations of multivariate Gaussian distributions, extended to infinitely many variables, $\boldsymbol{x}$. Thus, GPs can be interpreted as modelling distributions of \textit{functions} over infinite input domain. Formally:

\begin{definition}
    A Gaussian Process is a collection of random variables, any finite number of which have a joint Gaussian distribution.
\end{definition}

A GP for a real-valued process $y = f(\boldsymbol{x})$ is fully specified by its mean function, $m(\boldsymbol{x}) = \mathbb{E}[f(\boldsymbol{x})]$, and its covariance function, $k(\boldsymbol{x}, \boldsymbol{x}^\prime) =  \mathbb{E}[(f(\boldsymbol{x}) - m(\boldsymbol{x}))(f(\boldsymbol{x}^\prime) - m(\boldsymbol{x}^\prime))]$. This allows us to express the GP as:
\[
f(\boldsymbol{x}) \sim \mathcal{GP}(m(\boldsymbol{x}), k(\boldsymbol{x},\boldsymbol{x}^\prime))
\]

Any finite collection of function values has a jointly normal distribution: 
\[
[f(\boldsymbol{x_1}), f(\boldsymbol{x_2}), ..., f(\boldsymbol{x_n})]^T \sim \mathcal{N}(\boldsymbol{\mu}, K)
\]
where $\boldsymbol{\mu} = m(\boldsymbol{x})$ is the mean function and the covariance matrix $K$ is an $n \times n$ matrix whose entries are populated by the covariance function: $K_{i,j} = k(\boldsymbol{x}_i, \boldsymbol{x}_j)$. According to \cite{rasmussen_2006_gaussian}, we may assume that $\boldsymbol \mu = \boldsymbol 0$ for simplicity's sake, unless it is necessary to model a mean that varies over time. Under this assumption, a GP is defined entirely by its covariance function.\\

\section{Covariance Function}{\label{subsection:covariance_function}}

The covariance function encodes assumptions about the functions to be modelled. We can think of the covariance function as a measure of \textit{similarity} between data points $\boldsymbol{x}$ and $\boldsymbol{x'}$. One basic assumption is that close inputs of $\boldsymbol{x}$ return similar target values $y$. Note that for $K$ to be a valid covariance matrix, it must be positive semidefinite, such that $K$ satisfies $\boldsymbol{v}^T K \boldsymbol{v} \geq 0$ for all vectors $\boldsymbol{v} \in \mathcal{R}^n$. In \hyperref[fig:kernels]{Figure \ref*{fig:kernels}}, we illustrate how various common covariance functions determine the characteristics of functions randomly sampled from corresponding GPs. 

\begin{figure}[H]
    \centering
    \includegraphics[width=1\textwidth]{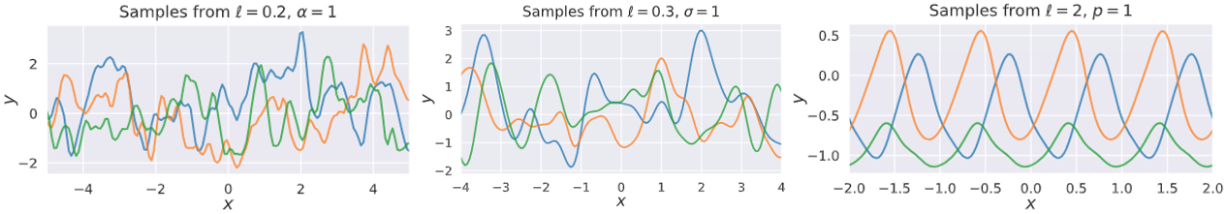}
    \caption{Here we present random samples drawn from three different GP models which have different underlying covariance functions. From left to right, we have Rational Quadratic, Exponentiated Quadratic and Periodic kernels \cite{roelants_2019_gaussian}. Hyperparameters are defined above each graph.}
    \label{fig:kernels}
\end{figure}

\section{Spectral Mixture Covariance Function}{\label{section:SM}
As outlined in {\cite{wilson_2013_gaussian}, the Spectral Mixture (SM) kernel is derived from a Mixture of Gaussians (MoG) in the frequency domain. This allows us to encode the underlying frequency components of a function, naturally providing a framework for modelling musical signals as superposed note sources. Furthermore, the SM kernel can approximate any covariance function to arbitrary precision by superposing sufficiently many weighted frequency components.\\

Importantly, the Spectral Mixture (SM) kernel is a stationary covariance function, which means it is a function of $\boldsymbol{x} - \boldsymbol{x'} =  \tau$. Thus it is invariant to translations of the input space. The inverse Fourier transform of the MoG allows us to find the time domain covariance function, with general form {\cite{wilson_2013_gaussian}: 

\begin{equation}{\label{equation:SM}}
 k( \tau) = \sum^Q_{q=1} w_q \prod_{p=1}^P \
\exp(-2\pi^2\tau_p^2 v_q^{(p)})\cos(2\pi\tau_p\mu_q^{(p)})
\end{equation}

where $P$ is the dimension of the inputs and $Q$ is the number of frequency components. The $q$-th Gaussian has weight $w_q$, and its $p$-th dimension has mean $\mu_q^{(p)}$ and variance $v_q^{(p)}$. In this work, we assume $P=1$ since audio is 1-dimensional, and we can omit references to $p$.

\section{Log Marginal Likelihood (LML)}
Assuming Gaussian noise $\sigma_n$ in our observed data $\boldsymbol{y}$ consisting of $n$ samples, we can derive the LML given our model hyperparameters\footnote{We refer to the parameters of the covariance function $k$ as `hyperparameters' to emphasise
that they are parameters of a non-parametric model.} $\boldsymbol{\theta}$ and data input vector $\boldsymbol{x}$:

\begin{equation}\label{eq:LML}
\log p(\boldsymbol{y}|\boldsymbol{x}, \boldsymbol{\theta}) = -\frac{1}{2}\boldsymbol{y}^T[ K+\sigma_n^2  I]^{-1}\boldsymbol{y} -\frac{1}{2} \log| K + \sigma_n^2  I| -\frac{n}{2} \log (2 \pi)
\end{equation}
Here, $ K$ is a function of $\boldsymbol x$ and $\boldsymbol \theta$ since $K_{i,j} = k({x}_i, {x}_j) = k(\tau)$ and $k$ is determined by $\boldsymbol \theta$. The first term of the LML is called the data fit term and the second is called the complexity term.

\part{III. Statistical Inference}
\chapter{High-level Approach}{\label{ch:high_level_approach}}
In this chapter, we summarise the overall framework of our score follower. This consists of four stages, of which the third stage is most substantial. Hence, we discuss that stage in the following two chapters. We also provide an overview of the methodology for implementation and evaluation. Finally, we cover the use of programming languages, data sources and software.

\section{Score Follower Framework}{\label{section:score_follower_framework}}
Our score follower comprises four main steps (as presented in \hyperref[fig:high_level]{Figure \ref*{fig:high_level}}):
\begin{enumerate}
    \item \textbf{Score Feature Extraction:} Important features from the score are extracted, including note onsets and pitches.   
    \item \textbf{Audioframe Extraction:} To perform real-time audio processing, we extract \textit{audioframes}, which are groups of contiguous audio samples, typically between 800 and 2000 samples long.
    \item \textbf{Score Following using Statistical Inference (Stages 1 and 2):} Score following is largely broken down into two statistical inference problems. In Stage 1: \textit{GP Model Specification}, we find the most probable underlying notes of isolated audioframes using a novel Gaussian Process (GP) model. In Stage 2: \textit{Real-Time Alignment}, we consider cumulative results from stage 1 and infer score location over time using Hidden Markov Models and state duration models.
    \item \textbf{Rendering of Score Position:} Using the results from step 3, score positions are displayed by a moving marker upon a digital score.  
\end{enumerate}

\begin{figure}[H]
    \centering
    \includegraphics[width = 0.7\textwidth]{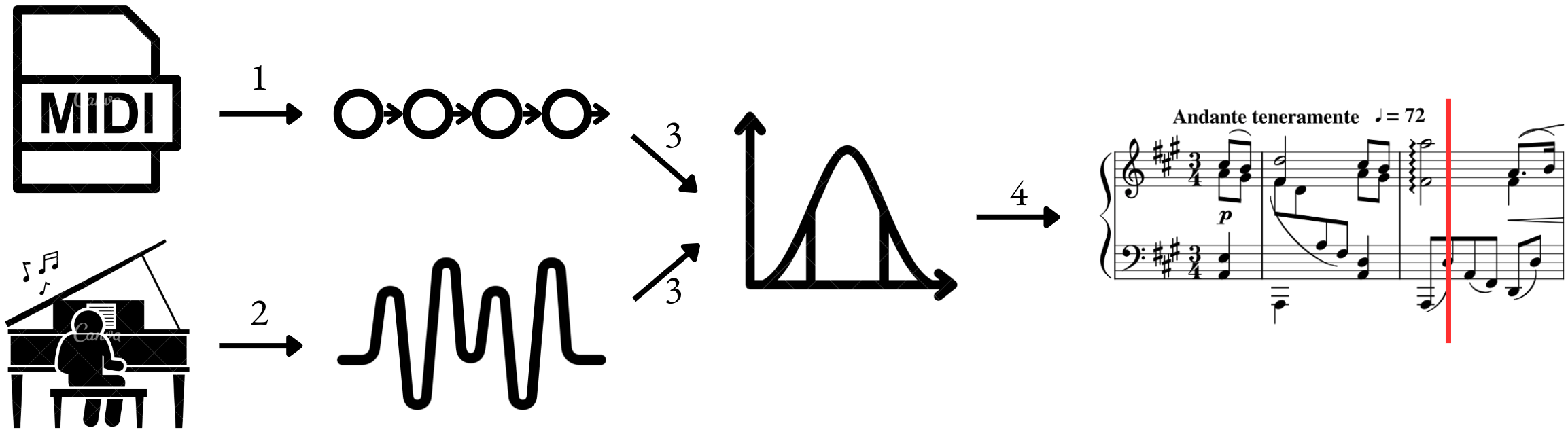}
    \caption{Diagram depicting the high-level score follower framework, where numbers represent the steps being completed.}
    \label{fig:high_level}
\end{figure}

\section{Overview of Methodology}
We began with step 3 and its two stages, since these constitute the bulk of the challenges. The detailed methods for these stages are respectively outlined in \hyperref[ch:model_selection]{chapter \ref*{ch:model_selection}} (Stage 1: \textit{GP Model Specification}) and \hyperref[ch:alignment]{chapter \ref*{ch:alignment}} (Stage 2: \textit{Real-Time Alignment}). We then approached steps 1, 2 and 4 in \hyperref[ch:implementation]{chapter \ref*{ch:implementation}}, which outlines the engineering details of how the final product brings all four steps together.

\subsection{Evaluation}
Because there is no widely accepted quantitative or objective method for evaluating score followers, we use a qualitative evaluation method based on the subjective evaluation of trained musicians. This is justified because score followers are ultimately to be used by musicians, and should meet their expectations.   

\section{Programming Language}
We chose Python since it is a high-level language with a rich ecosystem for Machine Learning tasks. Packages such as \verb|numpy| contain a significant amount of optimised code in C, which can mitigate the performance reductions resulting from the high-level, interpreted nature of Python. Though there exist Python packages that implement GPs, such as \verb|GPy|, \verb|GPyTorch|, and \verb|GPFlow|, all of our novel GP models were implemented without these packages to allow for flexibility. \verb|Jupyter| notebooks were used as an interactive development environment, especially during the early stages of modelling and hyperparameter optimisation.

\section{Sources}
\textbf{Kunstder Fugue:} MIDI files were downloaded from the website \textit{Kunstderfugue},\footnote{\href{https://www.kunstderfuge.com/}{https://www.kunstderfuge.com/}} which is a large resource for Classical Music in MIDI format. 

\textbf{Musescore:} MIDI and MusicXML files were downloaded from \textit{Musescore},\footnote{\href{https://musescore.com/sheetmusic?text=bach\%20fugue\%201}{https://musescore.com/sheetmusic?text=bach\%20fugue\%201}} a free sheet music website. 

\textbf{YouTube:} Experimental piano recordings were made by the author, and other recordings were collected on \textit{YouTube}.\footnote{\href{https://www.youtube.com/}{https://www.youtube.com/}}

\section{Software}
\textbf{Audacity:} All performance recordings made by the author were recorded and exported using \textit{Audacity},\footnote{\href{https://www.audacityteam.org/}{https://www.audacityteam.org/}} a widely used free program for recording and editing audio. 

\textbf{Flippy Qualitative TestBench:} The open source tool \textit{Flippy Qualitative Testbench}\footnote{\href{https://github.com/flippy-fyp/flippy-qualitative-testbench/tree/main}{https://github.com/flippy-fyp/flippy-qualitative-testbench/tree/main}} was used to render our score follower and evaluate our results. 
\newtcolorbox{lemmabox}[2][]{%
    colback=gray!20, 
    colframe=gray!60, 
    fonttitle=\bfseries, 
    title=#2, #1, 
    sharp corners, 
    boxrule=0.5mm, 
}
\chapter{Stage 1: GP Model Specification}{\label{ch:model_selection}}
 In this chapter we develop and optimise a Gaussian Process (GP) model for piano audio signals. This model gives us a log marginal likelihood (LML) function, which forms the basis for making predictions about underlying notes. The GP uses a Spectral Mixture (SM) covariance function, which we design through analysis of audio signals and optimise using both theoretical understanding of acoustics and empirical methods. In describing our methods, we discuss design decisions and trade-offs which affect the success of our score follower.      

\section{Aims and Requirements}{\label{section:aims}}
The primary goal of this chapter is to develop a log marginal likelihood (LML) function that estimates the notes present in an \textit{audioframe}. An audioframe is a group of contiguous audio samples, ranging in length between 800 and 2000 samples. Given a list of fundamental frequencies $\boldsymbol{f}$, the LML function should take an audioframe and output a log likelihood which:
\begin{itemize}
    \item is globally maximised when $\boldsymbol{f}$ is equal to the true underlying fundamental frequencies;
    \item is invariant to changes in other properties of the recording;
    \item is robust to the challenges mentioned in \hyperref[section:challenges]{section \ref*{section:challenges}};
    \item is efficient and stable to compute (since we are developing a real-time score follower).
\end{itemize}

\section{Motivation}
Gaussian Processes (GPs) offer a Bayesian framework for computing likelihoods over functions. They can be fine-tuned via their covariance functions to effectively capture the structured characteristics of audio signals (refer to \hyperref[ch:Gaussian Processes]{chapter \ref*{ch:Gaussian Processes}}).  \\

As concluded in \hyperref[ch:problem_definition]{chapter \ref*{ch:problem_definition}}, our objective is to ensure that the model accurately represents the harmonic structure of piano signals. As motivated in \hyperref[section:SM]{section \ref*{section:SM}}, we employ the Spectral Mixture (SM) kernel, which lends itself to being defined in the frequency domain, allowing us to feature engineer harmonic structure into the model. Moreover, the SM kernel incorporates a frequency hyperparameter, facilitating a simple interface for adjusting the model for different notes. Since we know all potential note combinations present in the score, we can systematically determine the LMLs of audioframes corresponding to the frequencies of those notes and determine the notes with maximal LML. By using the SM kernel, we assume that our GPs are stationary, meaning that they possess the same statistical properties throughout each audioframe. This is not strictly true due to decay and other noise transience effects, but it is a reasonable simplifying assumption since our audioframes are short in duration (800 to 2000 samples at $44.1$ kHz $\approx$ 18 to 45 ms). 

\section{Method}

\subsection{Signal Analysis}{\label{subsection:signal_analysis}}
We first took numerous recordings of different piano notes and chords at a sampling rate of $f_s = 44.1$ kHz and extracted various audioframes of 1000 and 2000 audio samples for analysis. See \hyperref[fig: example_audio_samples]{Figure \ref*{fig: example_audio_samples}} for one example of these recordings and an extracted audioframe.

\begin{figure}[H]
  \centering
  \begin{subfigure}[b]{0.45\textwidth}
    \centering
    \includegraphics[width=0.75\textwidth]{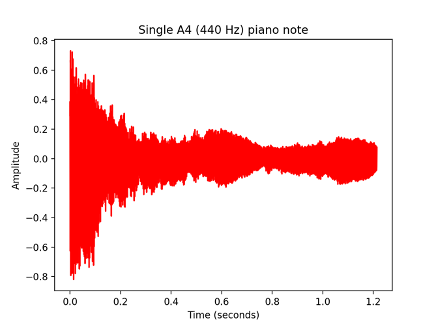}
    \caption{Time-amplitude graph of a recording of a single piano A4 note, zoomed out to show amplitude over a 1.2-second window containing dozens of audioframes.}
    \label{fig:zoom_out}
  \end{subfigure}
  \hfill
  \begin{subfigure}[b]{0.45\textwidth}
    \centering
    \includegraphics[width=0.75\textwidth]{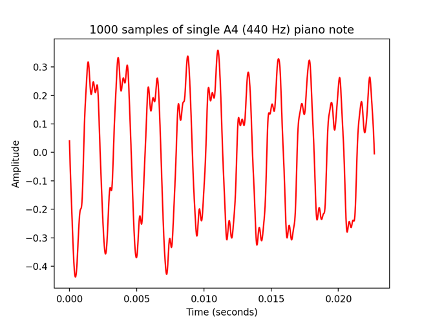}
    \caption{Time-amplitude graph of a single audioframe (1000 audio samples) extracted from \ref{fig:zoom_out}. Note the periodic but non-sinusoidal waveform.}
    \label{fig:zoom_in}
  \end{subfigure}
  \caption{Two time-amplitude graphs from a recording of the note A4 ($f_0 = 440$ Hz) on the piano.}
  \label{fig: example_audio_samples}
\end{figure}

Although \hyperref[fig:zoom_in]{Figure \ref*{fig:zoom_in}} represents a single A4 note with fundamental frequency 440 Hz, the time domain waveform is clearly not a pure sinusoid. Instead, the waveform demonstrates the natural interference patterns that arise from \textit{overtones}. These overtones are largely responsible for the unique \gls{timbre}s of different instruments (see \hyperref[ch:music_preliminaries]{chapter \ref*{ch:music_preliminaries}}). \\

We then plotted the magnitudes of the frequency and power spectra, using a Hanning window to avoid spectral leakage. As seen in \hyperref[fig: frequency_domain]{Figure \ref*{fig: frequency_domain}}, the harmonic content of a single note displays an accentuated peak at $f_0$, as well as several tapered peaks situated at the harmonics of the fundamental frequency, which are integer multiples of $f_0$. The relative magnitudes of these harmonics form the spectral `envelope' which we aim to model (see \hyperref[ch:music_preliminaries]{chapter \ref*{ch:music_preliminaries}}).  We also examined chords and found that their spectra were the superpositions of the underlying frequency spectra of each single note source.

\begin{figure}[H]
  \centering
  \begin{subfigure}[b]{0.45\textwidth}
    \centering
    \includegraphics[width=0.85\textwidth]{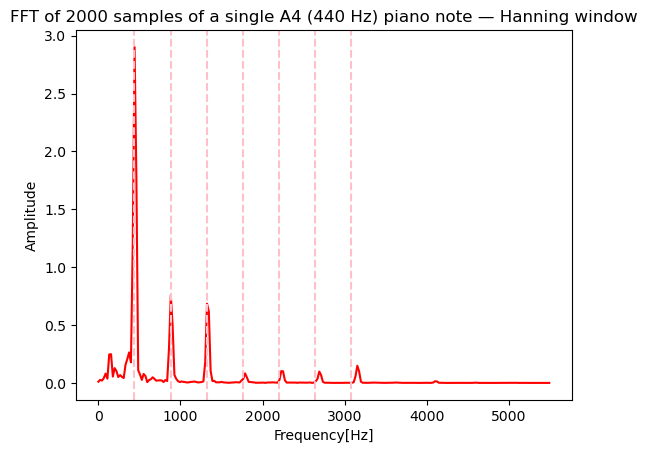}
    \caption{Frequency spectrum.}
    \label{fig:FFT}
  \end{subfigure}
  \hfill
  \begin{subfigure}[b]{0.45\textwidth}
    \centering
    \includegraphics[width=0.85\textwidth]{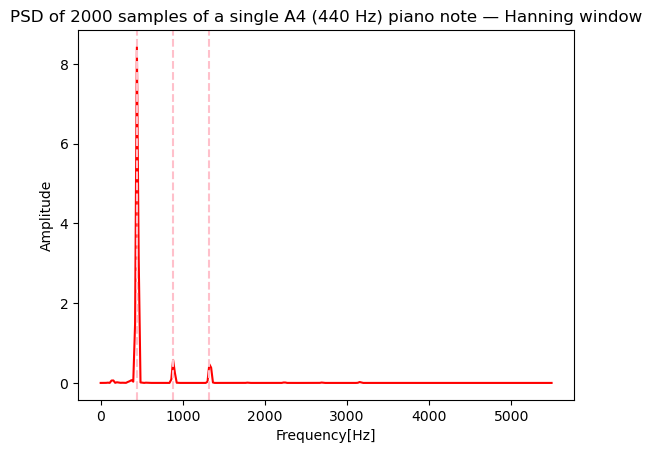}
    \caption{Power spectrum.}
    \label{fig:PSD}
  \end{subfigure}
  \caption{Examples of audio frequency spectra in a 2000-sample audioframe of a single A4 ($f_0 = 440$ Hz) piano note using a Hanning window. The overlaid pink dashed lines indicate the fundamental frequency and its integer multiples (harmonics).}
  \label{fig: frequency_domain}
\end{figure}

\subsection{Modelling}{\label{subsection:modelling}}
We initially model the above power spectrum as a Mixture of Gaussians:\footnote{We work in terms frequency $f$ since this representation of frequency lends itself better to musical notes.}

\begin{equation}{\label{equation:frequency}}
S(f) = \sum_{q=1}^Q w_q \sum_{m=1}^M \frac{E_m}{2} \left[ \phi(f; m f_q,\sigma_f) + \phi(f; -m f_q,\sigma_f) \right]
\end{equation}

Here, $Q$ represents the number of distinct note sources present, so $Q=1$ for single notes and $Q\geq 2$ for chords. Hence, $f_q$ represents the $q$-th note source's fundamental frequency. $M$ denotes the total number of harmonics that we model (including the fundamental), so the mean of the Gaussian for the $m$-th harmonic is $mf_q $. Thus, each of the peaks in the total power spectrum is represented by a Gaussian $\phi(f; m f_q,\sigma_f) = \frac{1}{\sigma_f\sqrt{2\pi}} \exp({-\frac{(f - m f_q)^2}{2\sigma_f^2}})$ with mean $m f_q$ and variance $\sigma_f^2$. The parameters $w_q$ and $E_m$ represent the relative weights assigned to the $q$-th note and $m$-th harmonic peaks, respectively. The values of $E_m$ are empirically determined to match the spectral envelope observed during analysis (see \hyperref[subsection:signal_analysis]{subsection \ref*{subsection:signal_analysis}}). Following \cite{godsill_2006_bayesian}, we approximate $E_m = \frac{1}{1 + Tm^v}$, where $T$ and $v$ are empirical constants.  

\subsubsection{Improving the Model}{\label{subsubsection:improving_model}}
We noticed several physical phenomena which compromised the model fit. For instance, we found a small unexpected peak at $50$ Hz which was due to a mains hum. This was resolved by simply adding a scaled Gaussian at $50$ Hz.\footnote{In America it was necessary to use a 60 Hz Gaussian.} Another phenomenon encountered was \textit{inharmonicity}, which is the tendency for overtones of a note to depart from whole multiples of $f_0$. This is a common musical phenomenon, occurring due to the physics of unforced oscillations of strings with finite tension \cite{murray_2021_musical}. In general, we noticed that harmonics consistently shifted upward, becoming sharper, with this effect being most pronounced at higher harmonics. 

\begin{figure}[H]
  \centering
  \begin{subfigure}[b]{0.48\textwidth}
    \centering
    \includegraphics[width=1\textwidth]{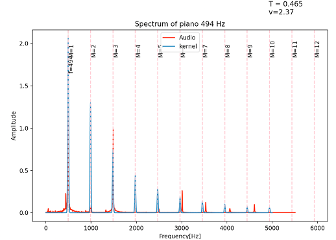}
    \caption{Audio spectra of a 2000 sample audio frame of a single B4 note (494 Hz) in red with the simple model overlaid in blue.}
    \label{fig:uncorrected}
  \end{subfigure}
  \hfill
  \begin{subfigure}[b]{0.48\textwidth}
    \centering
    \includegraphics[width=1\textwidth]{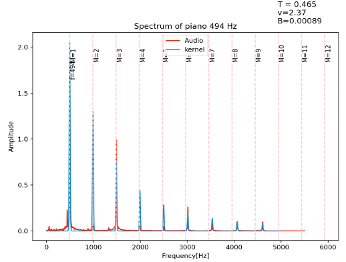}
    \caption{Same as (a), except kernel has been modified to account for inharmonicity.}
    \label{fig:corrected}
  \end{subfigure}
  \caption{Example of inharmonicity effects and how the multiplicative factor $b_{m,f_q}$ improves model fit. Values of $\boldsymbol{B}$ are determined in \hyperref[subsubsection:inharmonicity]{subsubsection \ref*{subsubsection:inharmonicity}}.}
  \label{fig:inharmonicity}
\end{figure}

 To account for inharmonicity, we use the approximation presented in \cite{Godsill2005BayesianCM}, which corrects the harmonic frequency $f_{m,q}$ of the $m$-th harmonic of the $q$-th note in the chord to $f_{m,q} = mf_q \sqrt{1 + B_{f_q}m^2} = m f_q b_{m,f_q}$ (for simplicity, let $b_{m,f_q}=\sqrt{1+B_{f_q}m^2}$). $B_{f}$ is an empirical constant for the note of the piano with fundamental frequency $f$, and all $B_f$ are stored in a vector $\boldsymbol{B}$. \hyperref[fig:inharmonicity]{Figure \ref*{fig:inharmonicity}} illustrates inharmonicity (\hyperref[fig:uncorrected]{Figure \ref*{fig:uncorrected}}) and our corrected harmonics (\hyperref[fig:corrected]{Figure \ref*{fig:corrected}}), where the disparity at higher frequencies between the true spectra (red) and the modelled spectra (blue) is eliminated by the multiplicative factor $b_{m,f_q}$.

\subsection{Covariance Function}{\label{subsection:covariance_function}}
Our model from \hyperref[subsection:modelling]{subsection \ref*{subsection:modelling}} forms the frequency spectrum of the SM kernel for our GPs. To obtain the time domain covariance function, we calculate the inverse Fourier transform of \hyperref[equation:frequency]{equation \ref*{equation:frequency}} (see \hyperref[appendix:iFT]{Appendix \ref*{appendix:iFT}}): 

\begin{equation}{\label{eq:k_tau}}
k(\tau) = e^{-2\pi^2\sigma_f^2 \tau^2} \sum_{q=1}^Q w_q \sum_{m=1}^M E_m \cos(2\pi m f_{q} b_{m,f_q}  \tau)
\end{equation}

where $\tau = x - x'$ represents the time between audio samples $x$ and $x'$ (since we are assuming a stationary covariance function; see \hyperref[section:SM]{section \ref*{section:SM}}). The other variables are hyperparameters and are defined in \hyperref[subsection:modelling]{subsection \ref*{subsection:modelling}}. This result is intuitive, comprising a quasi-periodic function of cosines multiplied by a negative exponential component. Each of these sinusoids has a frequency that depends on the underlying hyperparameters, corresponding to the frequency of each harmonic present in an audio wave. Due to linearity of the Fourier transform, the scalars $w_q$ and $E_m$, corresponding to the relative energy of each note source and defining the harmonic envelope respectively, are the same as in the frequency domain. The negative exponential component characterises the relationship between different audio samples as a function of $\tau$ and $\sigma_f$. Therefore, $\sigma_f$ represents a characteristic `inverse length scale', which determines the scale at which the function values have dependency on each other. \\

Thus far, we have seven hyperparameters: $\boldsymbol{\theta} = [\boldsymbol{f}, M,\sigma_f, \boldsymbol{w}, T, v, \boldsymbol{B}]$, recalling that $Q$ is the dimension of $\boldsymbol{f}$ and $E_m=\frac{1}{1+Tm^v}$ (see \hyperref[subsection:modelling]{subsection \ref*{subsection:modelling}} for definitions). In \hyperref[fig:covariance_functions]{Figure \ref*{fig:covariance_functions}} we plot several graphs demonstrating the effect of different hyperparameter combinations on our covariance function. In \hyperref[fig:random_samples]{Figure \ref*{fig:random_samples}}, we also plot three randomly drawn GP samples from these covariance functions with varying hyperparameters. These samples visually resemble the quasi-periodic audio signals observed in our initial analysis of piano note audio signals (see \hyperref[fig:zoom_in]{Figure \ref*{fig:zoom_in}}). Extending these random samples to seconds allowed us to successfully synthesise some musical sounds, which worked particularly well for the viola. The piano was more challenging to imitate, owing to its time-varying temporal envelope.

\begin{figure}
    \centering
    \includegraphics[width=0.96\textwidth]{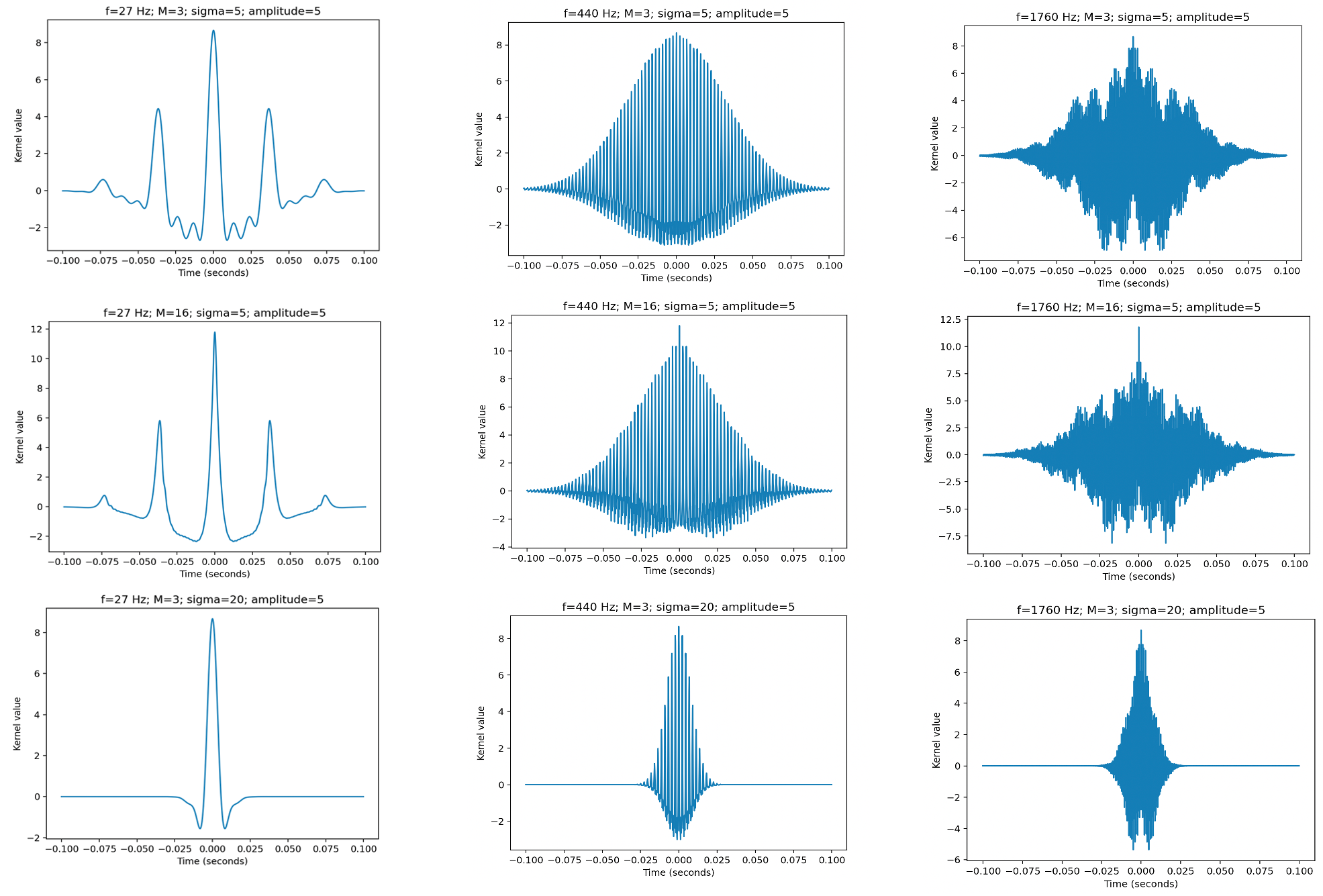}
    \caption{Illustration of the covariance function defined in \hyperref[subsection:covariance_function]{subsection \ref*{subsection:covariance_function}} and the effects of varying certain hyperparameters. The horizontal axis represents $\tau$, the time between two audio samples, and the vertical axis represents the corresponding covariance function values, intuitively the amount of dependence between two audio samples separated by $\tau$. In the left-to-right direction we have increasing fundamental frequency, which can be seen by the increasing number of periods. In the top row, we have $M=3$ and $\sigma_f = 5$. In the middle row, $M$ has increased to 16, as captured by the greater number of high-frequency fluctuations between the primary peaks, corresponding to the high frequency harmonics. In the top left, the three small peaks between the prominent peaks (corresponding to $M=3$) have been smoothed out in the middle left by 16 barely visible peaks which interfere with each other. Meanwhilst, in the middle and right column, the increase in $M$ results in more high frequency oscillations creating a more jagged waveform. Finally, in the bottom row we have increased $\sigma_f$ to 20. Increasing $\sigma_f$ decreases the length scale, which decreases the maximum $\tau$ at which there exists non-negligble dependencies between audio samples. This can be observed by the narrowness of the bottom row graphs.}
    \label{fig:covariance_functions}
\end{figure}

\begin{figure}
    \centering
    \includegraphics[width=0.96\textwidth]{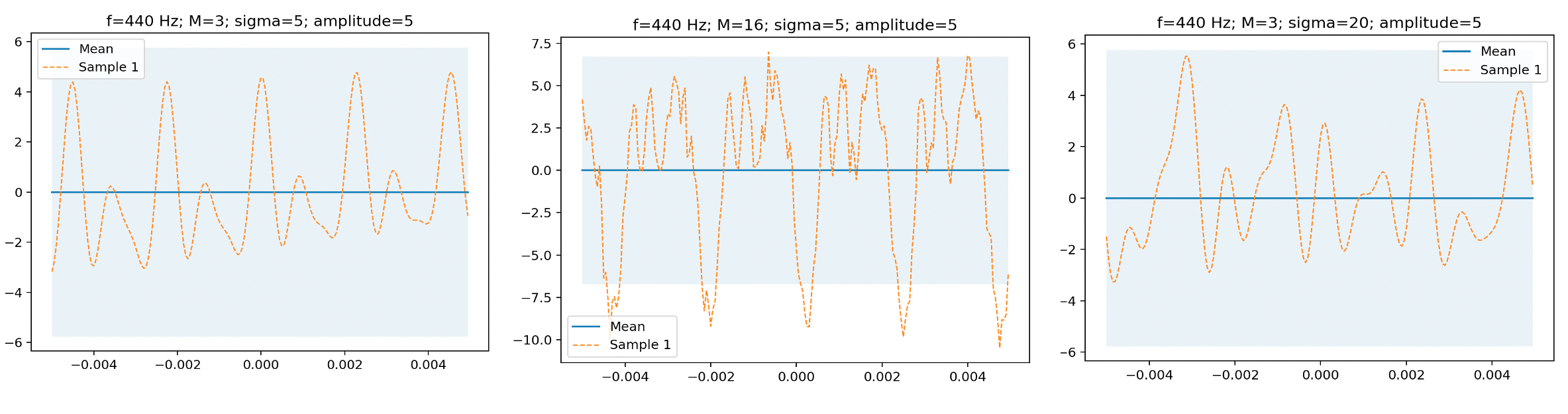}
    \caption{Three randomly drawn GP samples using the covariance function defined in \hyperref[subsection:covariance_function]{subsection \ref*{subsection:covariance_function}}. All three sub-figures have fundamental frequency 440 Hz, but differ in $M$ and $\sigma_f$. Increasing $M$ (centre vs. left) increases the complexity of the waveform due to more high-frequency components. Increasing $\sigma_f$ (right vs. left) decreases non-local dependencies and thus increases irregularity across the signal.      }
    \label{fig:random_samples}
\end{figure}

\subsection{Log Marginal Likelihood}{\label{subsection:LML}}
As detailed in \hyperref[ch:Gaussian Processes]{chapter \ref*{ch:Gaussian Processes}}, we utilise the log marginal likelihood (LML) to compare probabilities of different notes. This is defined in \hyperref[eq:LML]{equation \ref*{eq:LML}} and repeated here (with $\ell$ replacing $n$): 

\[ \log p(\boldsymbol{y}|\boldsymbol{x}, \boldsymbol{\theta}) = -\frac{1}{2}\boldsymbol{y}^T[K+\sigma_n^2 I]^{-1}\boldsymbol{y} -\frac{1}{2} \log|K + \sigma_n^2 I| -\frac{\ell}{2} \log (2 \pi) \]

where $\boldsymbol{y}$ is our audioframe containing $\ell$ samples, $\boldsymbol{x}$ contains the time offset of each audio sample, $\boldsymbol{\theta}$ is our hyperparameters, and $K$ is our covariance matrix, which is populated by our covariance function: $K_{i,j} = k(x_i, x_j)$. Since we know all possible note combinations from the score, we can compare all the LMLs for a given audioframe $\boldsymbol{y}$ given the $K$s computed from all the different corresponding frequency combinations (i.e. states). Note that this LML function includes $\sigma_n$ as an eighth hyperparameter representing Gaussian noise. Although we assume no noise in our GP model, we need to include noise when using the GP model on real-life data due to the inevitability of noise. It is also essential to add $\sigma^2_nI$ to $K$, since this mitigates stabilisation issues when computing the inverse of $K$, especially when $K$ is near singular. We leave $\sigma_n$ to be set by the user at runtime.

\subsection{Implementation of the LML Function}
We reduced calculation time by vectorising and using a ``sum of angles'' trick for the construction of $K$. Additionally we found that $\ell$, the number of audio samples in an audioframe, could be reduced to $800$ without compromising results. The LML requires calculating the inverse of the covariance matrix, $K^{-1}$, which has time complexity $\mathcal{O}(\ell^3)$ and may cause stability issues at small $\sigma_n$. To address these problems, we proceed by using a Cholesky factorisation (which also has complexity $\mathcal{O}(\ell^3)$, but is generally much faster) to reformulate the LML equation.  

\subsubsection{Cholesky Factorisation for the LML Function}
Since $K$ is symmetric positive definite (see \hyperref[subsection:covariance_function]{subsection \ref*{subsection:covariance_function}}), we can calculate $K^{-1}$ using a Cholesky factorisation, which is both faster to compute (the Cholesky factor takes  time $\frac {\ell^3} 6$) and highly numerically stable \cite{rasmussen_2006_gaussian}. We take a brief detour to derive this, since it is key to our stable implementation of the LML function. 

\begin{lemmabox}{Cholesky Factorisation}
\footnotesize
Let $A$ be a Hermitian, positive-definite matrix, $\boldsymbol{b}$ be a vector, and suppose $A\boldsymbol{x}=\boldsymbol{b}$. Factorise $A=LL^T$ for a lower triangular matrix $L$. Then: 

\[  A\mathbf{x} = \mathbf{b} \rightarrow LL^T\mathbf{x} = \mathbf{b} \rightarrow LL^T\mathbf{x} = L\mathbf{y} \rightarrow L^T\mathbf{x} = \mathbf{y} \]

If we solve for $L\boldsymbol{y} = \boldsymbol{b}$ and then solve for $\boldsymbol{x}$ in $L^{T}\boldsymbol{x} = \boldsymbol{y}$, we will have solved for $\boldsymbol{x}$ in our original $A\boldsymbol{x} = \boldsymbol{b}$ equation. Now, let the notation $A \backslash \boldsymbol{b}$ denote the vector $\boldsymbol{x}$ which solves $A\boldsymbol{x} = \boldsymbol{b}$. Then $\boldsymbol{x} = L^{T} \backslash (L \backslash \boldsymbol{b})$.  
\qed
\end{lemmabox}

To approximate the data fit term in \hyperref[eq:LML]{equation \ref*{eq:LML}}, we introduce a parameter $\boldsymbol{\alpha}$ such that $\boldsymbol{\alpha} = L^{T} \backslash (L \backslash \boldsymbol{y})$. Now, let $(K+\sigma^2_n I) = A = LL^T$. Hence, \[
-\frac{1}{2} \mathbf{y}^\top \left( K + \sigma_n^2 \mathbf{I} \right)^{-1} \mathbf{y} = -\frac{1}{2} \mathbf{y}^\top \boldsymbol{\alpha}
\]
For the complexity term of the LML, we can efficiently calculate the determinant: $
\det(K + \sigma_n^2 \mathbf{I}) = \det(LL^\top) = \det(L) \det(L^\top)$. Here, $\det(L) = \prod_{i=1}^\ell L_{i,i}$ because $L$ is a lower triangular matrix. Hence, 
\[\log \det (K + \sigma_n^2 \mathbf{I}) = \log \det (L^2) \\
= \log ( \left[ \prod_{i=1}^\ell L_{i,i} \right]^2 ) \\
=  2 \sum_{i=1}^\ell \log L_{i,i}
\]
Thus we have an efficient implementation of a stable LML function \cite{rasmussen_2006_gaussian}:
\[
\log p(\boldsymbol{y} | \boldsymbol{x}, \boldsymbol{\theta}) = -\frac{1}{2} \mathbf{y}^\top \boldsymbol{\alpha} - \sum_{i=1}^\ell \log L_{i,i} - \frac{\ell}{2} \log 2\pi
\]

\section{Hyperparameter Selection}


We aim to identify the true underlying notes of a single audioframe by comparing the values of the LML function over different possible fundamental frequencies $\boldsymbol{f}$ (i.e., states). Specifically, we infer that the true notes are those which correspond to the $\boldsymbol{f}$ that gives the highest LML. However, as established in \hyperref[subsection:covariance_function]{subsections \ref*{subsection:covariance_function}} and \hyperref[subsection:LML]{\ref*{subsection:LML}}, the LML function depends not only on $\boldsymbol{f}$, but also on seven other hyperparameters: $ \sigma_n, M, \sigma_f, \boldsymbol{w}, T, v$, and $\boldsymbol{B}$. Importantly, there are complex non-linear relationships between these hyperparameters, which means that the LML responds to \textit{combinations} of hyperparameters. Thus, if the LML for $\boldsymbol f_1$ is greater than the LML for $\boldsymbol f_2$, this could either be because $\boldsymbol f_1$ is closer to the true underlying notes, or because non-uniform interactions between $\boldsymbol f_1$ and other hyperparameters give $\boldsymbol f_1$ a higher LML despite $\boldsymbol f_1$ being \textit{farther} from the true notes. Hence, the optimal hyperparameters always maximise LML for the \textit{true} $\boldsymbol f$, avoiding the latter case above. However, finding these hyperparameters would require optimising over a highly non-convex 7-dimensional surface. This challenge is a paradigm case of the limitations of SM kernels. As Simpson, Lalchand and Rasmussen postulate, ``the key limitation in the SM kernel’s performance lies not in its stationarity or expressivity, but in the optimisation procedure''  \cite{simpson_2020_marginalised}.\\ 

Interestingly, \cite{simpson_2020_marginalised} proposes stochastic techniques like nested sampling and Monte Carlo methods as computationally tractable procedures to optimise hyperparameters. However, given that our goal is not necessarily true global optimality but rather a \textit{typical} set of hyperparameters that are sufficient for score following, we use our prior investigation of the hyperparameters and their physical bases to determine initial values and then empirically investigate LML graphs, using a mixture of local optimisation and tuning to obtain adequate values. In the subsequent subsections, we discuss the methods for choosing each of the hyperparameters. We begin with hyperparameters with clear physical interpretations, which makes them more straightforward to estimate. We then hold these as constant as we optimise the remaining hyperparameters. Note that we defer selection of $\sigma_n$ to the user, since this value is context-dependent. Even during this selection stage, we significantly varied $\sigma_n$ (ranging between 0.0001 and 10) to account for changing noise levels in different recordings. 

\subsubsection{$M$: Number of Harmonics}
$M$ represents the number of harmonics. Theoretically, $M$ could be very large since the harmonic series is infinite. However, this introduces a trade-off with computational complexity, which gives us reason to approximate the spectral envelope using a smaller value of $M$. As seen in \hyperref[subsection:signal_analysis]{subsection \ref*{subsection:signal_analysis}}, each subsequent harmonic has a lower energy, and visual inspection showed that harmonics above the 10th seem negligible. Therefore, we constrained our optimisation for $M$ ranging between 7 and 14, plotting graphs of LML vs. $M$ for inspection. In \hyperref[fig:M]{Figure \ref*{fig:M}} we show these graphs for two typical audioframes. As observed, we found that across all frequencies, the LML increases monotonically with $M$, but plateaus past a threshold of about 9. Hence, we set $M=9$ as an acceptable tradeoff for all frequencies.
 
 \begin{figure}[H]
  \centering
  \begin{subfigure}[b]{0.45\textwidth}
    \centering
    \includegraphics[width=0.75\textwidth]{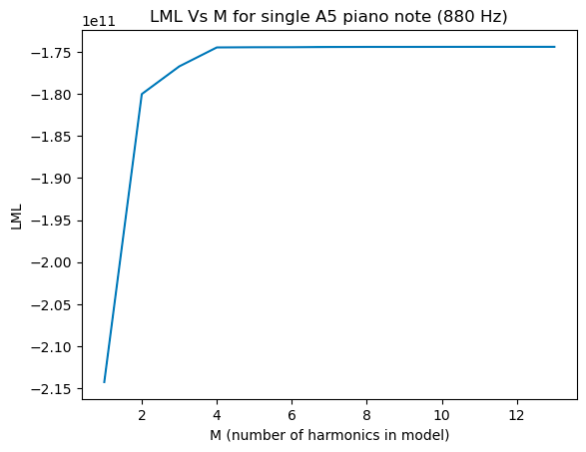}
    \caption{LML vs. $M$ for a single A5 note (880 Hz).}
    \label{fig:M_1}
  \end{subfigure}
  \hfill
  \begin{subfigure}[b]{0.45\textwidth}
    \centering
    \includegraphics[width=0.75\textwidth]{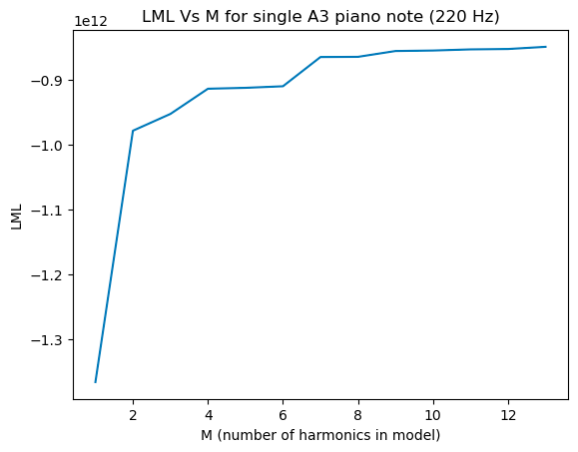}
    \caption{LML vs. $M$ for a single A3 note (220 Hz).}
    \label{fig:M_2}
  \end{subfigure}
  \caption{Graphs of LML against $M$ for two piano notes 2 octaves apart.}
  \label{fig:M}
\end{figure}
 
\subsubsection{$T$ and $v$: Spectral Envelope}
$T$ and $v$ determine the overall shape of the spectral envelope as the relative weights of the $m$-th harmonic is defined by $E_m=\frac{1}{1+Tm^v}$. To obtain a starting point for joint optimisation of $T$ and $v$, we used visual inspection of typical audioframe frequency spectra, overlaying the covariance function in the frequency domain (\hyperref[equation:frequency]{equation \ref*{equation:frequency}}) and systemically testing $T$ and $v$ values. Once we had obtained a reasonable estimate, we used gradient-based methods to find a jointly optimal pair over several typical audioframes using \verb|scipy.optimize|. Since the general shape of the spectra remained largely the same across frequencies, we settled for the optimised pair of $T = 0.465, v =2.37$ for all frequencies.

\subsubsection{$\boldsymbol{B}$: Inharmonicity Constants}{\label{subsubsection:inharmonicity}}
The inharmonicity constant $B_{f}$ depends on the behaviour of the piano strings for the note with fundamental frequency $f$. Piano strings vary in stiffness (see \hyperref[subsubsection:improving_model]{subsection \ref*{subsubsection:improving_model}}) not only over different $f$ but also over different pianos. Hence, it is difficult to generalise these values. For the piano used in this project, a dictionary was constructed to assign optimal $B_f$ values to all 88 keys. This was achieved by limited-memory BFGS, which is a quasi-Newton optimisation algorithm designed for large scale optimisation, again using \verb|scipy.optimize|.

\subsubsection{$\sigma_f$: Inverse Length Scale}
$\sigma_f$ is the standard deviation of the Gaussians in the frequency spectra (see \hyperref[subsection:modelling]{subsection \ref*{subsection:modelling}}). Thus, we can think of $\sigma_f$ as the spread of energy around the mean frequency, so higher $\sigma_f$ increases tolerance for deviation from expected frequency. In the time domain, $\sigma_f$ represents the inverse length scale, which defines the minimum time difference at which audio samples have negligible dependency (see \hyperref[fig:covariance_functions]{Figure \ref*{fig:covariance_functions}}). We began by estimating $\sigma_f$ by approximating the standard deviation of the visible peaks in our previous power spectra. This yielded $0.01$ as an order-of-magnitude starting point. Then, for audioframes of varying $\boldsymbol{f}$, we plotted LML vs. fundamental frequency for a logarithmic range of $\sigma_f$ centring around $0.01$, examining the graphs for prominent global maxima at the true fundamental frequency (see \hyperref[fig:sigma]{Figure \ref*{fig:sigma}}). Across all frequencies, we found that the smaller $\sigma_f$, the more prominent the LML peak at the true frequency. For $\sigma_f > 0.01$, there was significant error in the LML and imprecise peaks, and even false peaks at lower frequencies. We conjecture that for high $\sigma_f$, a wide range of low notes have high LML because low notes have many overtones, which allows overfitting if the tolerance for frequency deviation is too high. Finally, we settled for $\sigma_f = 0.005$.

\begin{figure}
  \centering
  \begin{subfigure}[b]{0.45\textwidth}
    \centering
    \includegraphics[width=0.75\textwidth]{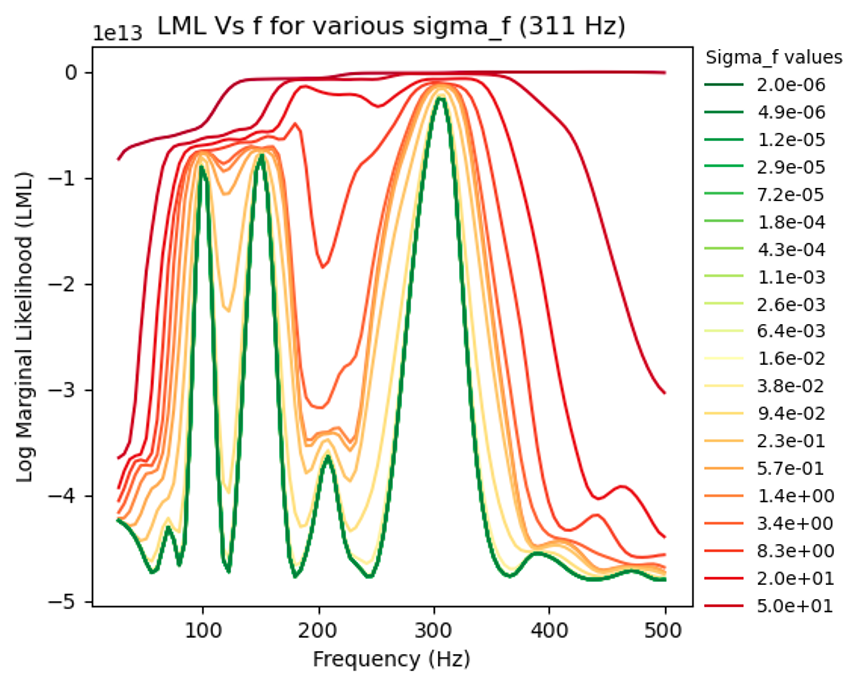}
    \caption{LML vs. frequency for a single D\#4 note (311 Hz).}
    \label{fig:sigma_1}
  \end{subfigure}
  \hfill
  \begin{subfigure}[b]{0.45\textwidth}
    \centering
    \includegraphics[width=0.75\textwidth]{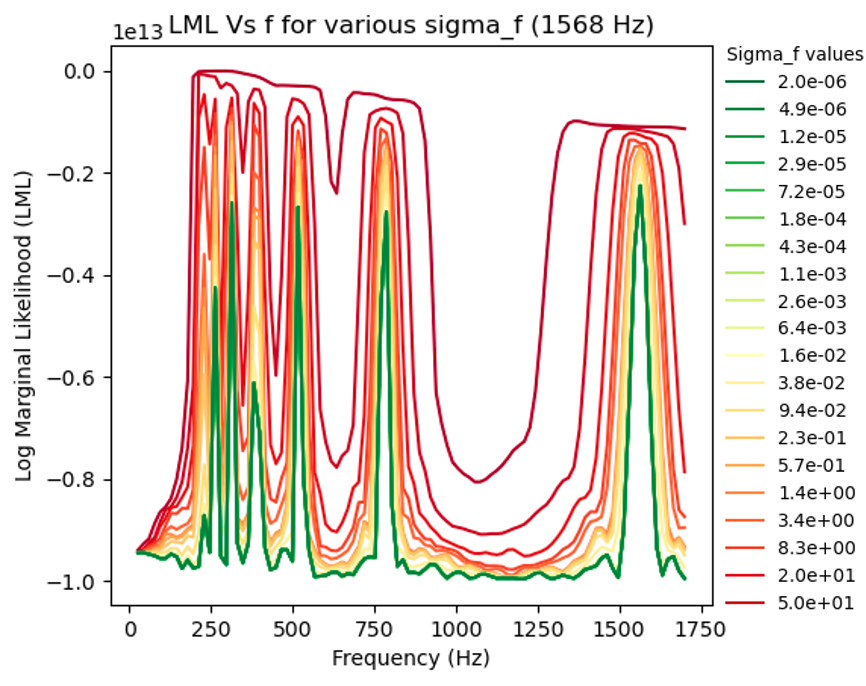}
    \caption{LML vs. frequency for a single G6 note (1568 Hz).}
    \label{fig:sigma_2}
  \end{subfigure}
  \caption{LML vs. frequency for various $\sigma_f$ (red is high, green is low) for notes over two octaves apart.}
  \label{fig:sigma}
\end{figure}

\subsubsection{$\boldsymbol{w}$: Relative Weights of Note Sources}{\label{subsubsection:amplitude}}
In audioframes containing multiple notes, the relative energies of those notes will almost always be non-uniform. This is due to physical phenomena like the properties of the hammers and strings on a piano, intentional musical techniques such as \textit{voicing}\footnote{Voicing is the technique of deliberately playing simultaneous notes at different loudness for emphasis.} and microphones with non-uniform frequency responses. Thus, an LML function that has $\boldsymbol{w}$ closer to the true relative energies of the note sources should yield higher a LML. We developed a method to estimate $\boldsymbol w$ using a least-squares approach to minimise the error between the true power spectrum and the modelled one. This involved a least-squares equation of the form $\boldsymbol{A} \boldsymbol{w} = \boldsymbol{\hat{b}}$, where each column of $\boldsymbol{A}$ is the power spectrum of a constituent note (modelled using our other hyperparameters) and $\boldsymbol{\hat{b}}$ is the $\boldsymbol w$-weighted sum of those power spectra, which estimates the resulting superposition of spectra. Minimising the squared residuals between $\boldsymbol{\hat{b}}$ and the true power spectrum of a multi-note audioframe allowed us to estimate the underlying $\boldsymbol w$ given $\boldsymbol{f}$. After estimating $\boldsymbol w$, we normalise it by dividing by $|w|_1$, so ultimately $|\boldsymbol w|_1=1$. \hyperref[fig:interval_amplitudes]{Figure \ref*{fig:interval_amplitudes}} illustrates the improvements of estimating $\boldsymbol w$ by plotting the modelled spectrum using estimated weights over a true spectrum. 

\begin{figure}[H]
  \centering
  \begin{subfigure}[b]{0.45\textwidth}
    \centering
    \includegraphics[width=0.95\textwidth]{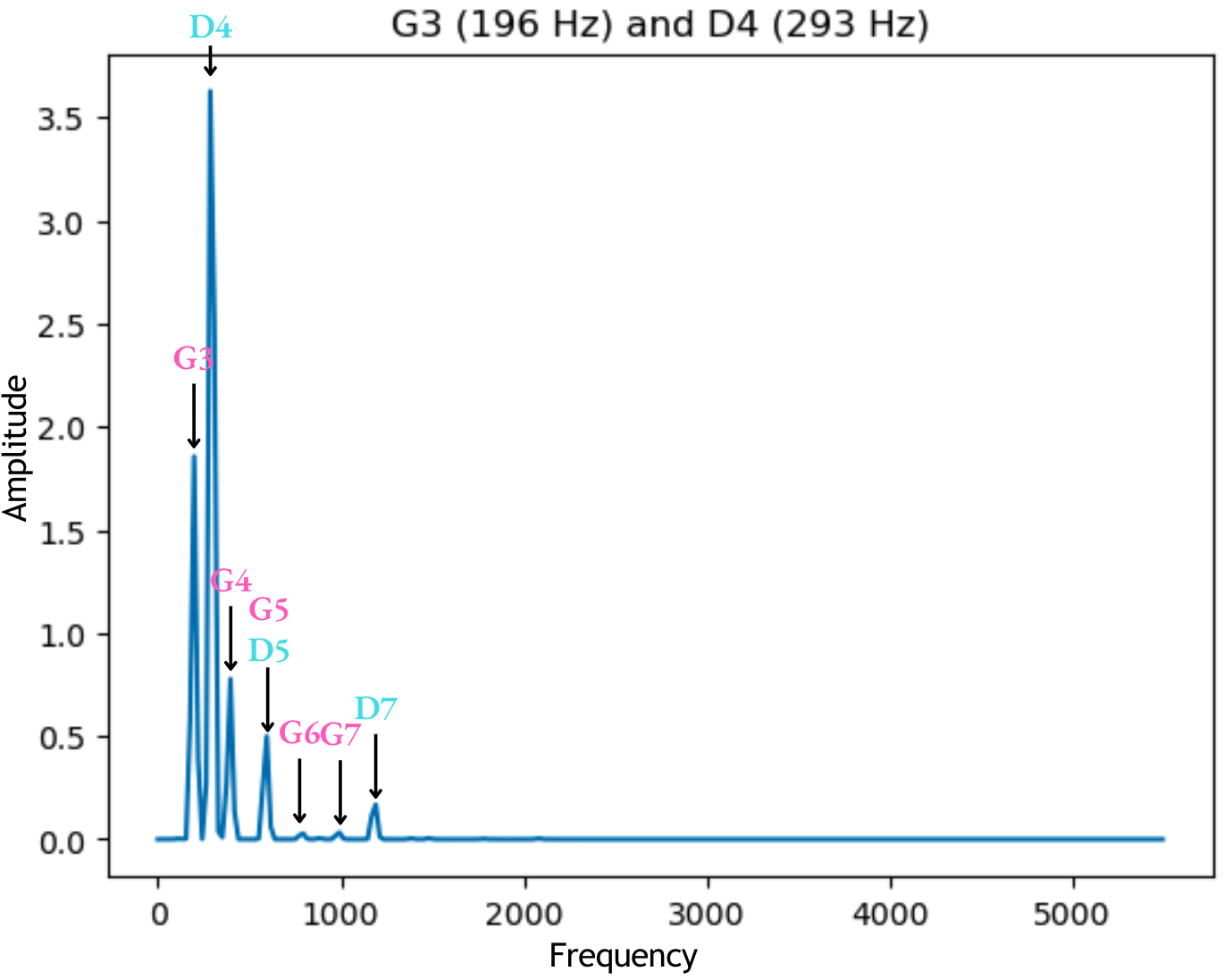}
    \caption{True power spectrum of note sources G3 and D4 with partials labelled in pink and turquoise respectively ($\boldsymbol{f} = [196, 283]$).}
    \label{fig:interval}
  \end{subfigure}
  \hfill
  \begin{subfigure}[b]{0.45\textwidth}
    \centering
    \includegraphics[width=0.95\textwidth]{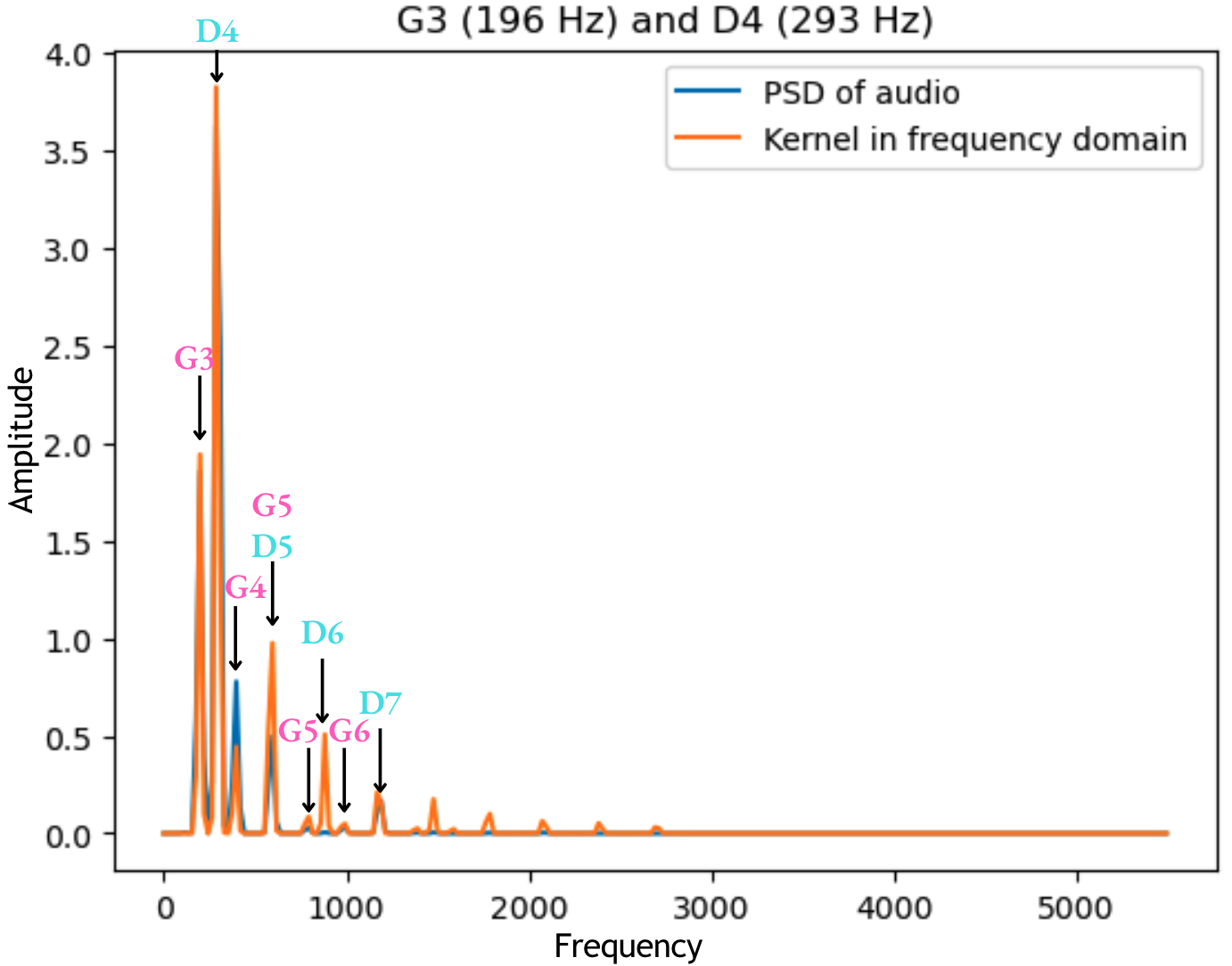}
    \caption{\hyperref[fig:interval]{Figure \ref*{fig:interval}} with power spectrum of modelled covariance function overlaid in orange, using amplitudes found via a least squares approach.}
    \label{fig:amplitude}
  \end{subfigure}
  \caption{Graphs of G3, D4 interval, exhibiting large relative amplitude difference (D4 is much louder) and how least squares can be used to estimate relative note source weights.}
  \label{fig:interval_amplitudes}
\end{figure}

 Although this model was closer to the true spectra, the improvements to the LML values were minimal and often inconsistent. For example, in the recording of a soft G3 and loud D4, the LML assuming uniform weighting (i.e. $\frac{\boldsymbol{w}}{{|\boldsymbol{w}|}}=[0.5,0.5]^T$) was 2398, whereas the LML using the normalised relative weights (i.e., $\frac{\boldsymbol{w}}{{|\boldsymbol{w}|}}= [0.216, 0.784]^T$) calculated from the least squares method gave us 2416. This is an insignificant increase compared to the changes in LML observed from changes in $\boldsymbol{f}$. Furthermore, the least squares calculations can only be performed once the audioframes are received, increasing runtime. Hence, we decided to simply assume $\boldsymbol w = \mathbf {\frac 1Q}$. Ultimately, this was of little consequence for score following, since our GP model is much more sensitive to notes having the wrong frequencies rather than the wrong amplitudes, due to the nature of the SM kernel used with a small value of $\sigma_f$. However, estimating $\boldsymbol w$ could be useful in other applications, such as sound synthesis.

 \section{Results and Discussion}
After tuning our hyperparameters, we finally have a fully defined GP model with a SM kernel, which can now be used to infer notes from audioframes. The key result from this chapter is shown in \hyperref[fig:LML]{Figure \ref*{fig:LML}}. The largest peak occurs at the true fundamental frequency of the audioframe, as indicated by the red dashed line. The other peaks correspond to the octaves below, where some of the partials line up. Note that the y-axis scale is logarithmic, implying that the seemingly small delta between adjacent peaks is in fact substantial, with the first difference in likelihood being a factor of $e^{326}$. Similar results hold across a range of notes and chords, with the exception of very similar chords which have many but not all notes in common and therefore share many harmonics.

\begin{figure}[H]
    \centering
    \includegraphics{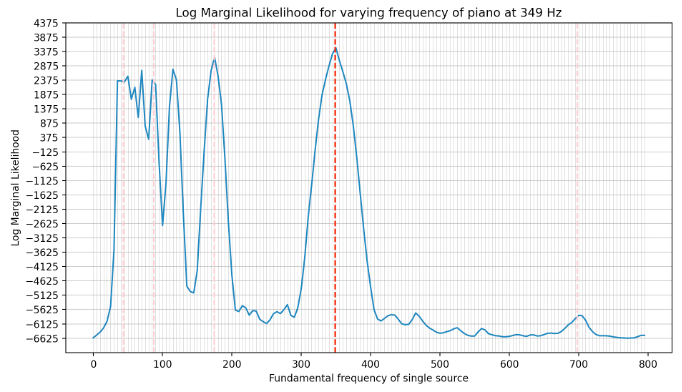}
    \caption{Final result illustrating the variation in LML for varying frequency of a single F4 piano note (349 Hz). Note the global peak at the true underlying frequency (red dashed lines). }
    \label{fig:LML}
\end{figure}

Overall, we have successfully met the requirements for this chapter as set out in \hyperref[section:aims]{section \ref*{section:aims}}. The results from the likelihood function produce a clear spike when $\boldsymbol{f}$ is the true underlying fundamental frequencies, and this is largely invariant to changes in the recording parameters. Due to our optimisations, the LML function is fairly efficient and stable to compute, as well as robust to the challenges mentioned in \hyperref[section:challenges]{section \ref*{section:challenges}}.



\chapter{Stage 2: Real-Time Alignment}{\label{ch:alignment}}
In this chapter, we implement a statistical algorithm to infer score location in real time from note likelihoods calculated using the log marginal likelihood (LML) function developed in \hyperref[ch:model_selection]{chapter \ref*{ch:model_selection}}. This algorithm uses a Hidden Markov Model (HMM), where audioframes (groups of contiguous audio samples) are the observed variables and notes in the score are the latent states. We use the note LML function as emission probabilities for the HMM, and we develop a state duration model for transition probabilities. This allows us to formulate a Maximum A Posteriori (MAP) estimate for the most likely sub-sequence of latent states ending at the current time. Finally, we solve for the MAP using a `Windowed' Viterbi algorithm, which efficiently returns the most likely current score position.


\section{Aims and Requirements}{\label{section:aims_and_reqs_Alignment}}
The primary goal of this chapter is to develop a statistical method for real-time score following building on the Gaussian Process (GP) model developed in \hyperref[ch:model_selection]{chapter \ref*{ch:model_selection}}. We require an online alignment algorithm which:
\begin{itemize}
    \item predicts the most probable state, given all audioframes that have been observed so far;
    \item meets the time-sensitive demands of a real-time application with no noticeable lag;
    \item is robust to the challenges mentioned in \hyperref[section:challenges]{section \ref*{section:challenges}}.
\end{itemize}
For the scope of this project, we assume that the pianist starts at the beginning of the piece and makes no significant deviations from the score.

\section{Motivation}
Hidden Markov Models (HMMs) are a natural choice to address the problem of score following, as musical signals can be modelled as a sequence of observations (audioframes) that depends upon a discrete latent state process (musical notes). If we can predict these latent states from the observations, we will have access to score position. Furthermore, HMMs are a highly suitable choice for this particular project since the log marginal likelihood (LML) function developed in \hyperref[ch:model_selection]{chapter \ref*{ch:model_selection}} provides the emission probabilities for this HMM. \\

Additionally, the use of transition probabilities in a HMM presents the opportunity to incorporate knowledge of the score's \gls{rhythm}. This is because rhythm provides information about when the next note is likely to occur, and this information can be encoded in the HMM's transition probabilities. Thus, we make use of both pitch and note duration, the two perceptual features established to be helpful for score following in \hyperref[subsection:score_influences]{subsection \ref*{subsection:score_influences}}.


\section{Hidden Markov Model}{\label{section:HMM}}
A Hidden Markov Model (HMM) is a tool for modelling probability distributions over sequences of observations that depend upon some latent (or `hidden') process. HMMs have been extensively used in many applications, such as speech recognition, gene sequencing, signal processing and, of course, score following (refer to \hyperref[section:literature_review]{section \ref*{section:literature_review}}). 

\subsection{Hidden Markov Model for Score Following }
Let $\boldsymbol{y}_n$ represent the $n$-th observed variable (i.e. the $n$-th of $N$ audioframes). Then, let $s_n \in \{1, \dots , K\}$ represent the latent states we would like to predict (i.e. the underlying notes), sorted by time in the score. $s_n$ can take one of $K$ discrete states, such that $s_n=k$ means that the latent state at audioframe $n$ is the $k$-th chord or note in the list of possible states. \hyperref[fig:HMM]{Figure \ref*{fig:HMM}} shows a representation of this HMM, where the grey nodes are latent states and the white nodes are observations. The sequence of latent states is a stochastic process, where transitions are constrained such that the system can only either stay in the same state (self-transition) or move to the next state (advance-transition). This is because we assume that the pianist makes no mistakes, so there is no skipping of states or backtracking. Moreover, we can assume a first-order HMM, since the transition probabilities do not depend on previous states.\footnote{Since we ultimately implement a state-duration model, the transition probabilities are determined by the current state and number of self-transitions \textit{d}, but not previous states. Thus, this model might most accurately be called a \textit{time-varying} first-order HMM.} Hence, this a left-to-right first-order Markov chain, which is illustrated in the state transition diagram in \hyperref[fig:latent]{Figure \ref*{fig:latent}}. 

\noindent 
\begin{minipage}{0.46\textwidth}
\centering
    \includegraphics[width=0.85\linewidth]{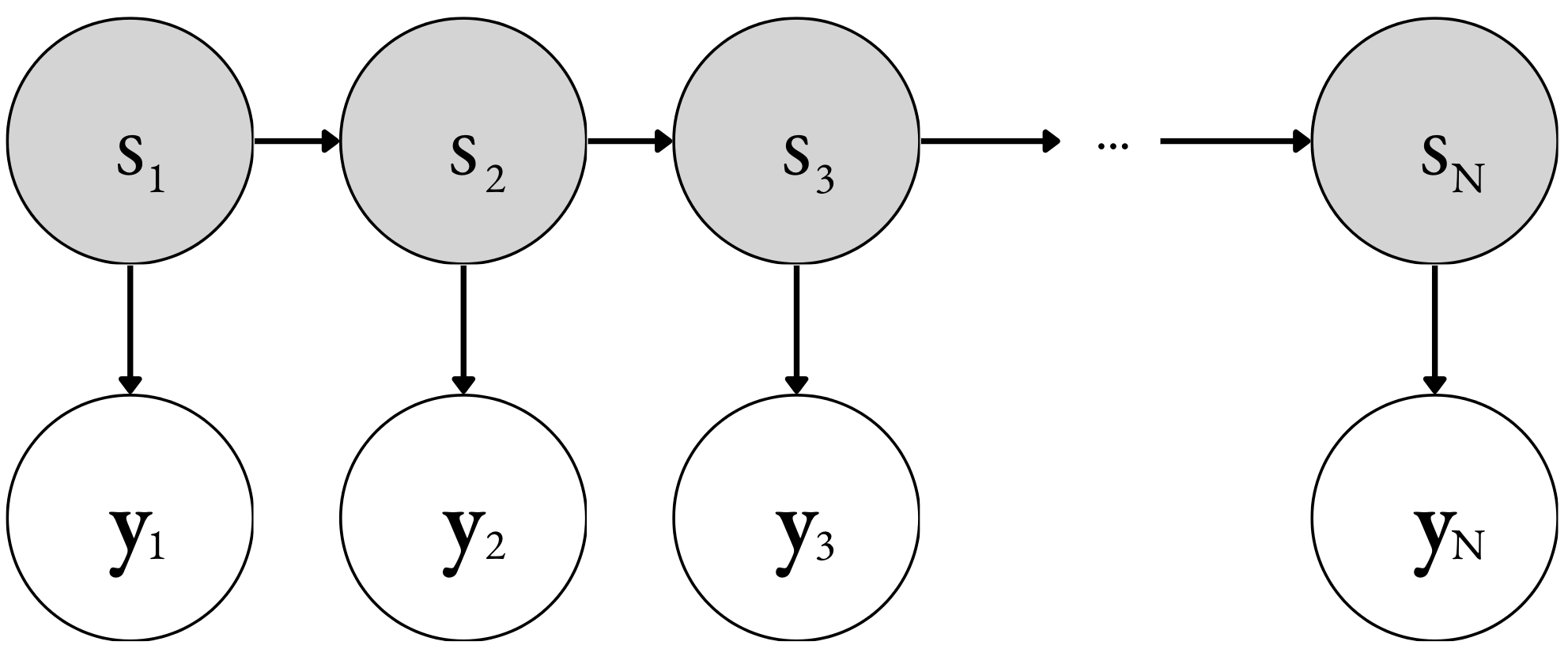}
    \captionof{figure}{Illustration of the HMM. Latent states $s_n$ are represented by grey nodes and the observed variables $\boldsymbol{y}_n$ are represented by white nodes.}
    \label{fig:HMM}
\end{minipage}%
\hfill 
\begin{minipage}{0.46\textwidth}
\centering
    \includegraphics[width=0.85\linewidth]{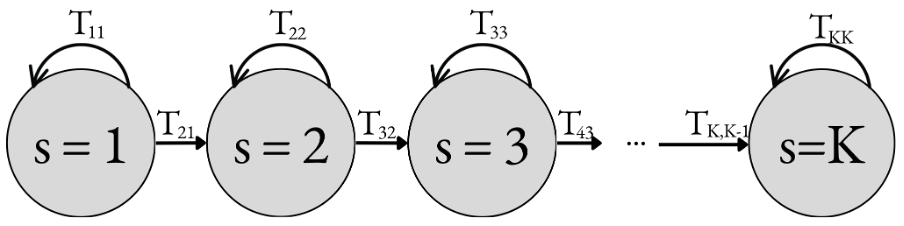}
    \captionof{figure}{Illustration of the left-to-right first-order state transition diagram for the latent states $s$ (represented as grey nodes). $T_{i,j}$ represents the transition probability from state $j$ to state $i$.}
    \label{fig:latent}
\end{minipage}\\

The joint distribution of the sequence of latent states and observations can be written as:

\begin{equation} \label{eq:joint_distribution}
P(\mathbf{y}_{1:N}, s_{1:N}) = P(s_1)P(\mathbf{y}_1|s_1) \prod_{n=2}^N P(s_n|s_{n-1})P(\mathbf{y}_n|s_n)
\end{equation}

 As we assume the pianist starts at the beginning of the piece, we set the initialisation probability to 1, such that $P(s_1) = P(s_1 = 1) = 1$.

\section{State Duration Model}{\label{section:state_duration_model}
HMMs contain a transition probability function, represented above as $P(s_n|s_{n-1})$. Unlike typical HMMs, whose transition probabilities can be specified by a constant transition matrix, in the HMM for our score follower, the transition probabilities depend on how long the current note has already been played for. This is because the longer a note has been sustained already, the higher the probability of an advance-transition, whilst a note that has just begun is most likely to continue (self-transition). Similar to note pitch information, we can extract note duration information from a MIDI. Hence, we can use estimates of state durations to approximate the likelihood of state transitions. This enables our model to account for not only pitch, but also \gls{rhythm}. We discuss our state duration model and our estimates of note durations, respectively, in the following two subsections. Note that our approach differs from the formal State Duration HMMs in \cite{Dewar_2012}. This presents an area for possible future work. 

\subsection{Geometric Distribution for State Transitions}
We use a geometric distribution (over non-negative support) to model the number of self-transitions, $Z$, before an advance-transition, such that $Z \sim Geo(p)$ for some probability $p$. The mean of this distribution is $\mathbb E [Z]$, the expected number of self-transitions before an advance-transition. Intuitively, for a given note, $\mathbb E [Z]$ equals the expected number of audioframes that fit in that note. $\mathbb E[Z]$ determines $p$, the probability of an advance transition in our geometric distribution, as $p = \frac{1}{1+\mathbb{E}[Z]}=1- \frac{\mathbb{E}[Z]}{1 + \mathbb{E}[Z]}$. \\

The CDF of the geometric distribution allows us to calculate the transition probabilities, given $d$, the number of self-transitions that have already occurred. If $d$ self-transitions have already occurred, the probability of another self-transition equals $P(Z>d) = 1-P(Z\leq d)$. Since we exclude all transitions except self- and advance-transitions, the probability of an advance-transition equals the geometric CDF, or $1-(1-p)^{d+1}$.\\

Substituting $p=1-\frac{\mathbb E[Z]}{1+\mathbb E[Z]}$, we can formulate our transition function between two consecutive states as follows:
\begin{equation}
P_d(s_n=i|s_{n-1}=j) = 
\begin{cases}
\big( \frac{\mathbb E[Z]}{1+ \mathbb E[Z]} \big )^{d+1} & i=j\\[1.5ex]
    1 - \big( \frac{\mathbb E[Z]}{1+ \mathbb E[Z]} \big )^{d+1} & i=j+1 \\[1.5ex]
    0 & \text{otherwise}
\end{cases}
\end{equation}

We can then use this transition function to populate elements of our transition matrix:\footnote{Here, $T^d$ does not refer to exponentiation, but rather the transition matrix $T$ after $d$ self-transitions.} $\boldsymbol{T}^d_{i,j} = P_d(s_n=i|s_{n-1}=j)$.

\subsection{Estimating $\mathbb E [Z]$}

A rudimentary approximation of $\mathbb E[Z]$ assumes that, on average, the pianist plays perfectly at the tempo specified in the MIDI. Then, the expected number of audioframes in a note is given by $\mathbb{E}[Z] = \verb|time_to_next| \times f_s$, where $\verb|time_to_next|$ is the length of this note in seconds (specified by the MIDI) and $f_s$ is the audioframe sampling frequency. In this method, $\mathbb E[Z]$ can be precomputed for every note state in the score.\\




However, the method above does not consider that performers almost always deviate in tempo from the one specified in the MIDI. The performer may continually change the tempo, both due to expression and error, so a good score follower should adjust to local fluctuations in tempo. Thus, we calculate a tempo conversion factor, \verb|conversion_rate|, which provides a local estimate of the number of audioframes per second of MIDI time (not real time). For each of the preceding notes, we know both the MIDI note length in seconds (\verb|time_to_next|), as well as the observed value of how many audioframes corresponded to that note ($Z$). Thus, we obtain the \verb|conversion_rate| by computing a moving average of $\frac {Z}{\texttt{time\_to\_next}}$ over the preceding $h$ notes (where we can adapt $h$ to suit the performance). This allows us to locally estimate $\mathbb{E}[Z] = \texttt{conversion\_rate}\times \texttt{time\_to\_next}$, using the current note's MIDI duration as \verb|time_to_next|. Although this method allows adjusting for tempo, it does not allow precalculating $\mathbb E[Z]$ for all notes. Nonetheless, since the moving average is not a computationally intense calculation, this does not compromise the the HMM's real-time performance.


\section{The Viterbi Algorithm}{\label{section:general_viterbi}}
Given an HMM, the Viterbi algorithm allows us to predict the most likely historical sequence of latent states (notes) given \textit{all} observations (audioframes). In its original form, the Viterbi algorithm is more suitable for score alignment than score following, since it requires all audioframes to optimise the sequence of notes (see \hyperref[subsection:score_following_v_alignment]{subsection \ref*{subsection:score_following_v_alignment}}). However, in our real-time application, we only have access to the first $n$ audioframes. Importantly, upon receiving a new audioframe, we cannot simply append a new predicted state to the previously predicted sequence (i.e., a greedy algorithm), since the new observation may alter our beliefs about past states. Due to the computational complexity of the general Viterbi algorithm, it is infeasible to re-execute the whole algorithm after each audioframe. \\

As outlined in \cite{rmek_2007_the}, there exist online Viterbi algorithms, including ones which are still guaranteed to produce optimal solutions, like the short-time Viterbi algorithm in \cite{bloit_2008_shorttime}. However, the short-time Viterbi algorithm does not provide predictions at \textit{every} observation (audioframe). Another potential solution could be to greedily use a `$k$-best' Viterbi algorithm, where we track the most probable $k$ paths to the current audioframe \cite{Brown2010DecodingHU}. However, given the sequential structure of music, the sparsity of our transition matrix, and the constraints imposed by the state duration model, we know that the $k$ most likely paths are likely close to one another, seldom differing by more than a certain number (\textit{window size}) of states. To take advantage of the structure of our HMM, we opt to implement a `Windowed' Viterbi algorithm, which is still suitable for online use. First, however, we examine the general Viterbi algorithm.

\subsection{The General Viterbi Algorithm}

To predict the sequence $s_{1:N}$ based on the observed data $\boldsymbol{y}_{1:N}$, we use the Maximum A Posteriori (MAP) estimate, which involves finding $s_{1:N}$ that maximises the posterior:
\[
\boldsymbol{s}^* = \underset{\boldsymbol{s}}{\operatorname{arg\,max}} \; P(s_{1:N}|\boldsymbol{y}_{1:N}) = 
 \underset{\boldsymbol{s}}{\operatorname{arg\,max}} \; \frac{P(\boldsymbol{y}_{1:N},s_{1:N})}{P(\boldsymbol{y}_{1:N})} =  \underset{\boldsymbol{s}}{\operatorname{arg\,max}} \; P(\boldsymbol{y}_{1:N},s_{1:N})
\]

Taking logarithms and substituting \hyperref[eq:joint_distribution]{equation \ref*{eq:joint_distribution}} gives:
\begin{equation}
    \boldsymbol{s}^* = \underset{\boldsymbol{s}}{\operatorname{arg\,max}} \bigg\{\log P(s_1) + \log P(\boldsymbol{y}_{1}|s_1) + \sum_{n=2}^N \big[
\log P(\boldsymbol{y}_{n}|s_n) + \log P(s_n|s_{n-1})\big] \bigg\}
\end{equation}
The first two terms are invariant to $s$, since we define $s_1=1$ (we always start at the beginning of the piece). The first term in the sum, $\log P(\boldsymbol{y}_{n}|s_n)$, is the log emission probability, which we obtain using the Gaussian Process (GP) LML function from \hyperref[subsection:LML]{subsection \ref*{subsection:LML}}. Since the GP LML takes hyperparameters $\boldsymbol{\theta}$ and not a state $s_n$, we encode $s_n$ into a corresponding $\boldsymbol{\theta}$ using the hyperparameter $\boldsymbol{f}$ (see \hyperref[subsection:covariance_function]{subsection \ref{subsection:covariance_function}}). The second term in the sum, $\log P(s_n|s_{n-1})$, is the log of the transition probability from \hyperref[section:state_duration_model]{section \ref{section:state_duration_model}}.\\

The Viterbi algorithm solves for $\boldsymbol{s^*}$ with dynamic programming: it uses a bottom-up approach to calculate the most likely sub-sequence until each audioframe. This involves a maximisation routine at each audioframe via an exhaustive search over all possible next states. We start at audioframe 1, which has a single trivial solution. Then, we build up a table of possible paths, calculating the highest probability subsequence to each state after each audioframe. Once the subproblem spans all $N$ audioframes, we have arrived at an optimal solution, and can complete a final traceback to calculate the entire optimal sequence $\boldsymbol{s^*}$. This algorithm has a computational complexity of $\mathcal{O}(K^2 N)$, where $K$ is the number of note states and $N$ is the number of audioframes. Refer to \cite{viterbi} for more details of the Viterbi algorithm.\\

As outlined in \hyperref[section:HMM]{section \ref*{section:HMM}}, our HMM is special in that it is a left-to-right first-order Markov chain. Hence, our Viterbi algorithm can be adapted to be computationally simpler than the general case, as the transition matrix $\boldsymbol{T^d}$ is sparse: $T^d_{i,j}=0$ unless $i-j \in \{ 0,1\}$. We show this in the trellis diagram in \hyperref[fig:trellis]{Figure \ref*{fig:trellis}}, which illustrates all possible sequences of states across the first five audioframes. This diagram shows that at each audioframe except the first, there are only two possible previous states. Nonetheless, the requirement that we provide predictions of score location in real-time requires us to further optimise the Viterbi algorithm.
\begin{figure}[H]
    \centering
    \includegraphics[width=0.45 \textwidth]{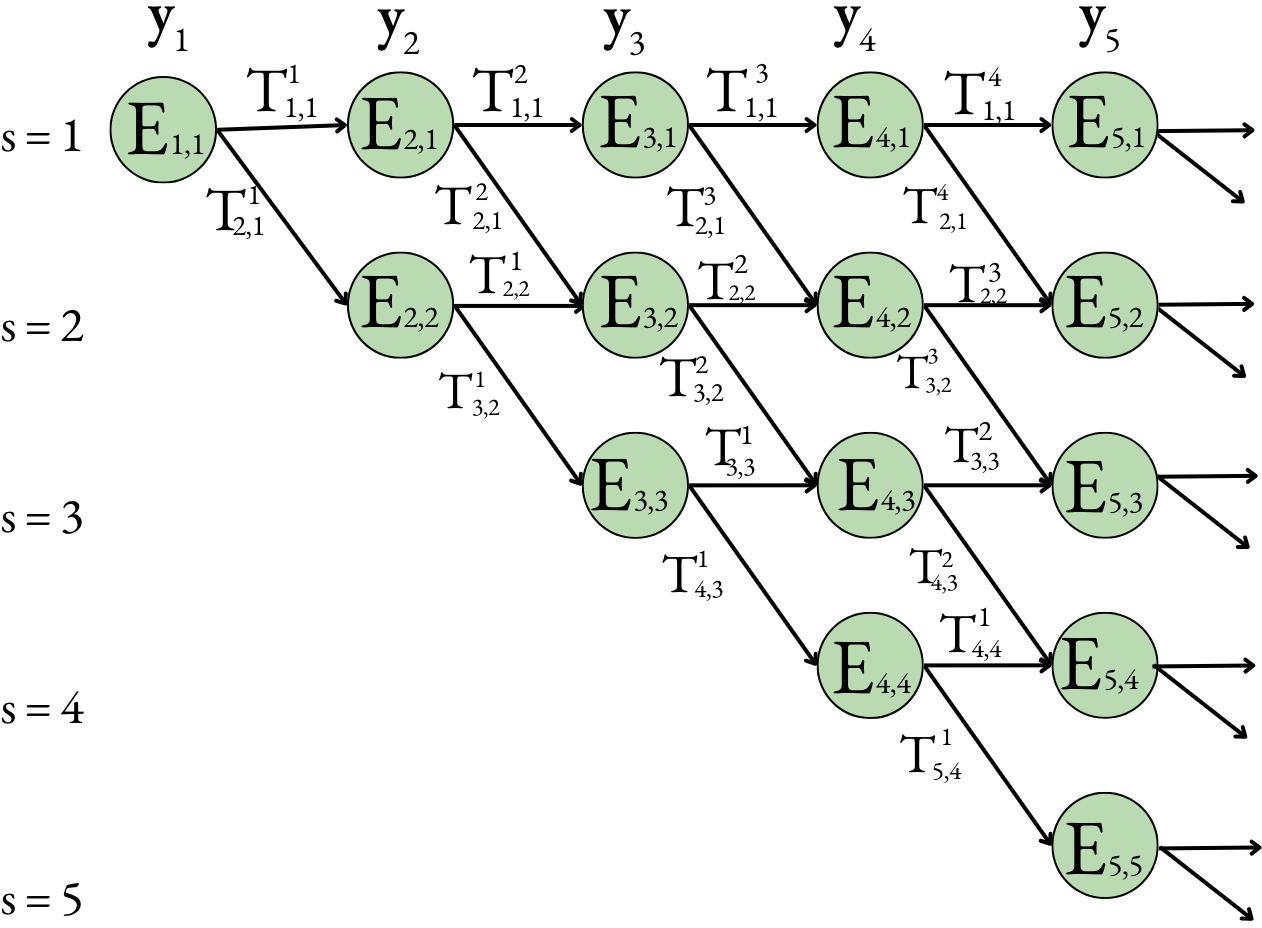}
    \caption{Trellis diagram of the set of possible paths during the first five audioframes of the Viterbi algorithm. Rows represent the different states $s=k$, and columns represent each audioframe $\boldsymbol{y}_n$. The values in the $n$-th column of the dynamic programming matrix, $\boldsymbol \Pi$, are defined by the total product of the path across the first $n$ audioframes (or alternatively the sum of the log probabilities), where $T^d_{i,j}$ represents the transition probabilities and $E_{i,j}$ represents the emission probabilities. Note that $T^d_{i,j}$ varies with $d$, the number of self transitions that have already occurred, as detailed in \hyperref[section:state_duration_model]{section \ref*{section:state_duration_model}}.  }
    \label{fig:trellis}
\end{figure}

\subsection{Adapting the Viterbi Algorithm}{\label{subsection:adjusting_viterbi}}

For the reasons introduced in \hyperref[section:general_viterbi]{section \ref*{section:general_viterbi}}, we design an approximately optimal `Windowed' Viterbi algorithm, in which we only examine states which sit within a \textit{window} of constant length $\ell$ that advances alongside the current note state, $s_n$. The trellis diagram in \hyperref[fig:windowing]{Figure \ref*{fig:windowing}} depicts the execution of this algorithm with a window length $\ell=6$ that shifts forward once its position within the window exceeds a threshold of $\varphi=4$. 

\begin{figure}[H]
  \centering
  \begin{subfigure}[b]{0.49\textwidth}
    \centering
    \includegraphics[width=0.8\textwidth]{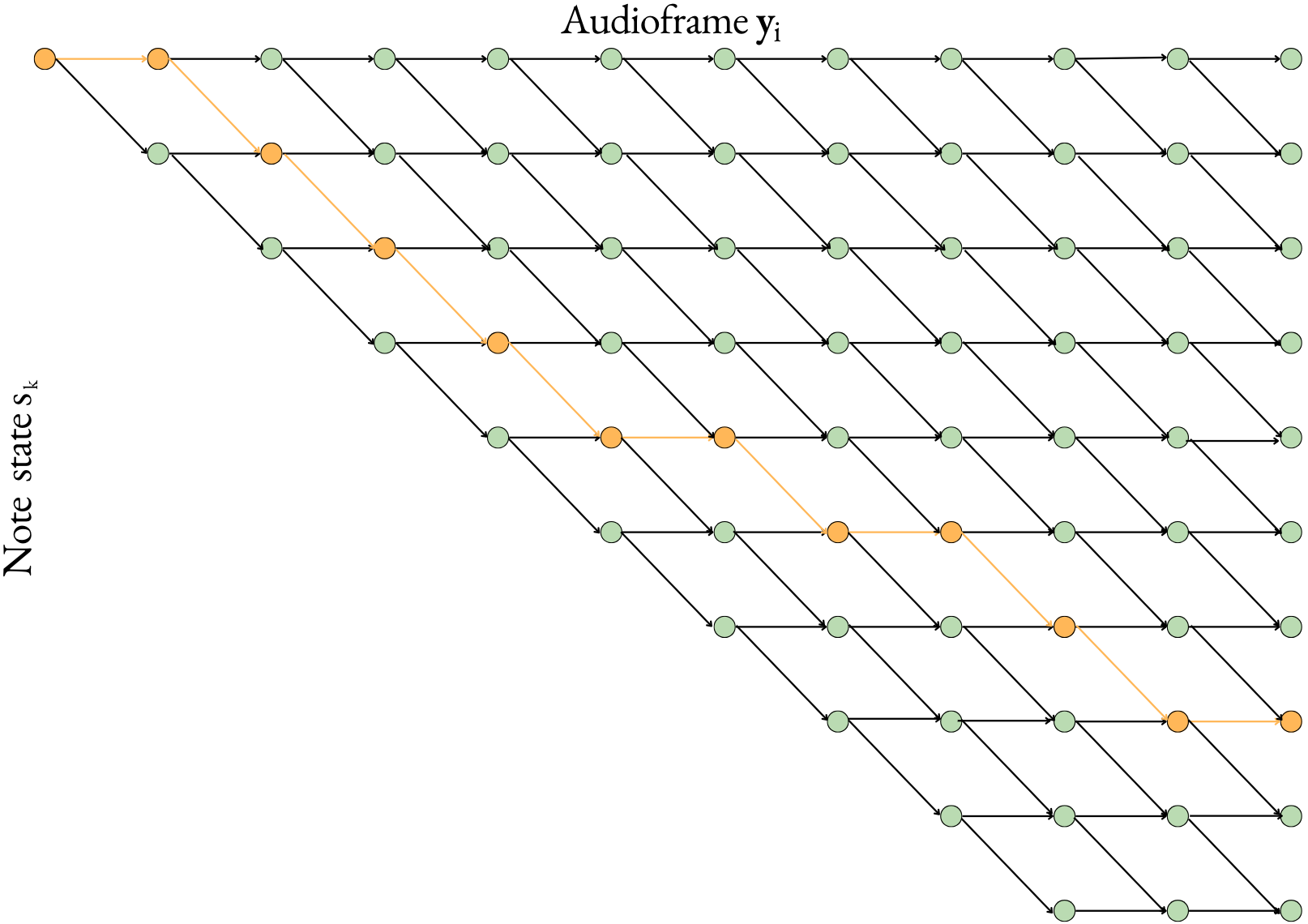}
    \caption{The trellis diagram of a general Viterbi algorithm for a left-to-right first-order Markov chain for a piece with 10 notes. We see that with each new audioframe, the number of states searched increases, until we reach the last state. }
    \label{fig:unpruned}
  \end{subfigure}
  \hfill
  \begin{subfigure}[b]{0.49\textwidth}
    \centering
    \includegraphics[width=0.8\textwidth]{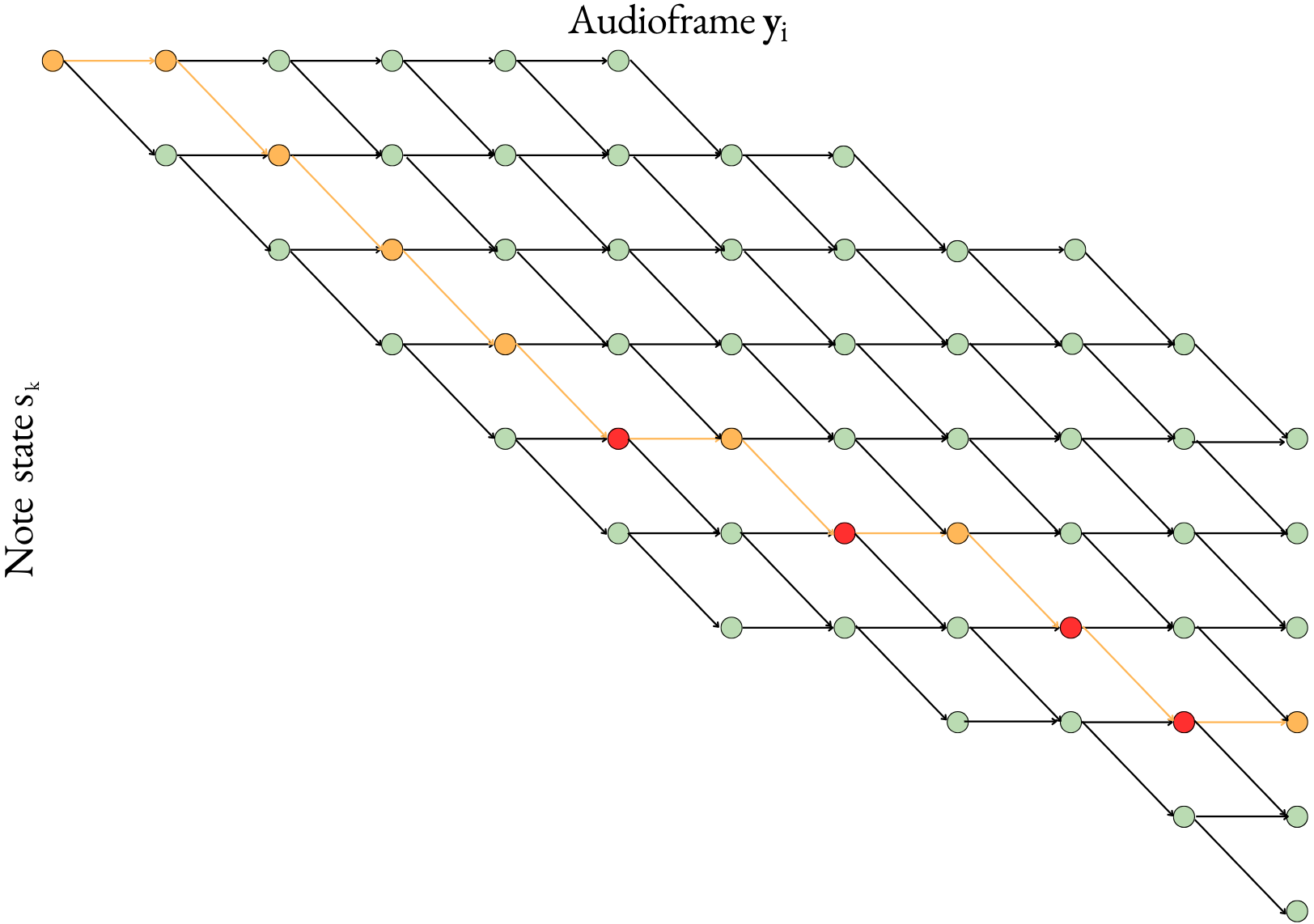}
    \caption{Trellis diagram of our Windowed Viterbi algorithm, where the \textit{window} size is 6 and the threshold position in the window is 4. The red nodes represent the nodes which cross the threshold and cause the next iteration to advance the window by one.}
    \label{fig:pruned}
  \end{subfigure}
  \caption{Two trellis diagrams illustrating the general and Windowed Viterbi algorithm.}
  \label{fig:windowing}
\end{figure}

The inner for loop of this algorithm populates the $n$-th column of the probability matrix $\boldsymbol{\Pi}$, where $\boldsymbol\Pi[i,j]$ is the probability of the most likely path to state $s_i$ at audioframe $\boldsymbol y_j$.

\begin{algorithm}
\footnotesize
\caption{$n$-th inner loop of Windowed Viterbi Algorithm}
\begin{algorithmic}

\Require audioframe $\boldsymbol{y}_{n}$, path probability matrix $\boldsymbol{\Pi}$, window start $w$, window length $\ell$,\\ \ \ \ \ \ \ \ \ \ \ \  window threshold $\varphi$, previous state $s_{n-1}$, transition probabilities $\boldsymbol T^d$
\For{$k \gets w$ to $w+ \ell$}
    \State $\boldsymbol{\Pi}[k, n] \gets \max \big \{\mathrm{LML}(\boldsymbol{y}_{n}|s_n=k) + T_{k,k}^{d} + \boldsymbol{\Pi}[k, n-1] \ \ ,\ \  \mathrm{LML}(\boldsymbol{y}_{n}|s_n=k) + T_{k-1,k}^{d} + \boldsymbol{\Pi}[k-1, n-1]\big \}$
\EndFor

\State $s_n \gets \mathrm{argmax}_k \big \{\boldsymbol{\Pi}[k, n]\big \}$
\If{$s_n > w + \varphi$}
    \State $w \gets s_n - \varphi$
\EndIf
\end{algorithmic}
\end{algorithm}

Owing to the high sampling frequency compared to the relatively slow note changes in real music, we would expect many more self-transitions than in this toy example, resulting in a more horizontal trajectory through the trellis. This is the reason the Windowed Viterbi algorithm's approximated solution is acceptable. In the unmodified Viterbi algorithm, since we have many self-transitions, probabilities outside the window will approach zero since we repeatedly multiply by near-zero probabilities. Hence, we can justify modelling the transition probabilities to and from these states as zero, effectively pruning all paths outside the window.

\section{Results and Discussion}{\label{section:stage_2_results}}
Testing the success of Stage 2: \textit{Real-Time Alignment} of the score follower involved recording short snippets of piano pieces, running the Windowed Viterbi algorithm, then overlaying the predicted states on a time-amplitude graph of the recordings. Visual inspection (and some manual time-mapping) was then used to analyse the correctness of these predictions. Note that this is a limited method for evaluation, since only pieces with visually obvious state transitions can be used.  Moreover, the method of manually checking each audioframe is tedious, highlighting the need for a live score position renderer for evaluation (see \hyperref[section:renderer]{section \ref*{section:renderer}}). \\

\hyperref[fig:intermediate_results]{Figure \ref*{fig:intermediate_results}} shows the key results from this section. We have used Bach's \textit{Prelude in C major BWV 846} because this is a \gls{monophonic} piece, so its states are visually easy to identify. For recordings where no or some \gls{sustain pedal} was used, the results were very successful, as presented in \hyperref[fig:no_pedal]{Figures \ref*{fig:no_pedal}} and \hyperref[fig:some_pedal]{\ref*{fig:some_pedal}} . However, use of very generous pedal caused results to become temperamental, as seen in the erroneous results of \hyperref[fig:loads_pedal]{Figure \ref*{fig:loads_pedal}}. \\

\begin{figure}[H]
  \centering
  \begin{subfigure}[b]{0.33\textwidth}
    \centering
    \includegraphics[width=1\textwidth]{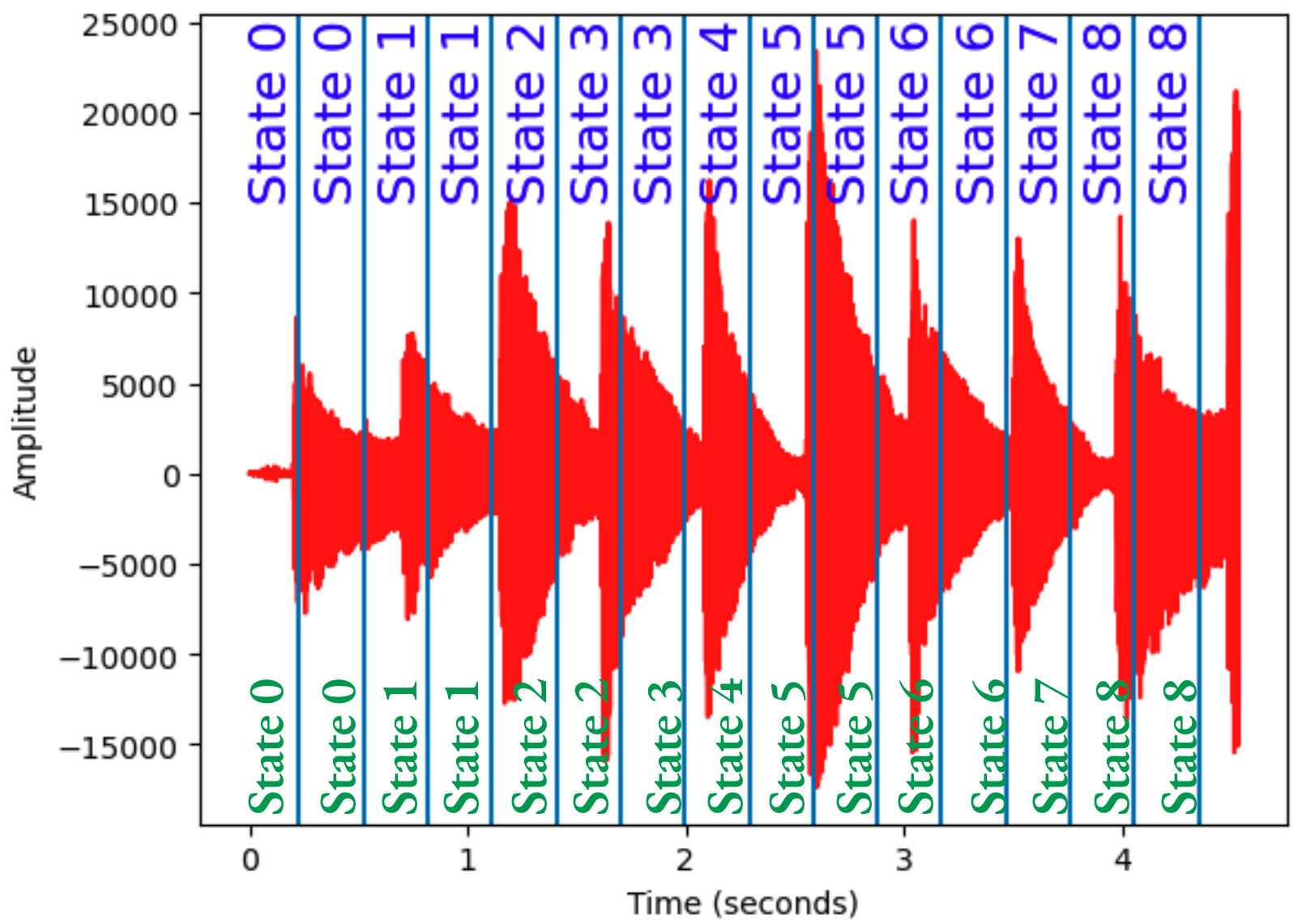}
    \caption{Recording with no pedal. All predictions are correct.}
    \label{fig:no_pedal}
  \end{subfigure}
  \hfill
  \begin{subfigure}[b]{0.32\textwidth}
    \centering
    \includegraphics[width=1\textwidth]{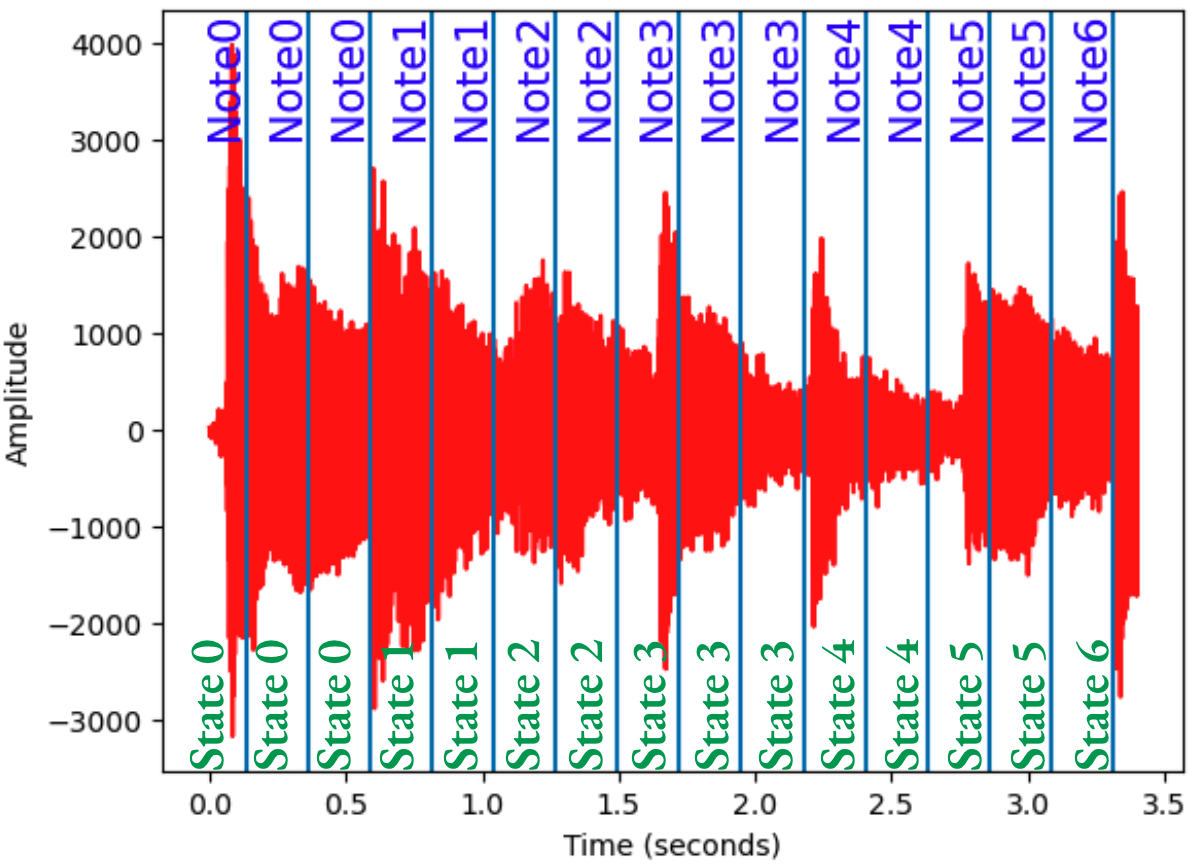}
    \caption{Recording with a realistic amount of pedal. All predictions are correct.}
    \label{fig:some_pedal}
  \end{subfigure}
  \hfill
  \begin{subfigure}[b]{0.33\textwidth}
    \centering
    \includegraphics[width=1\textwidth]{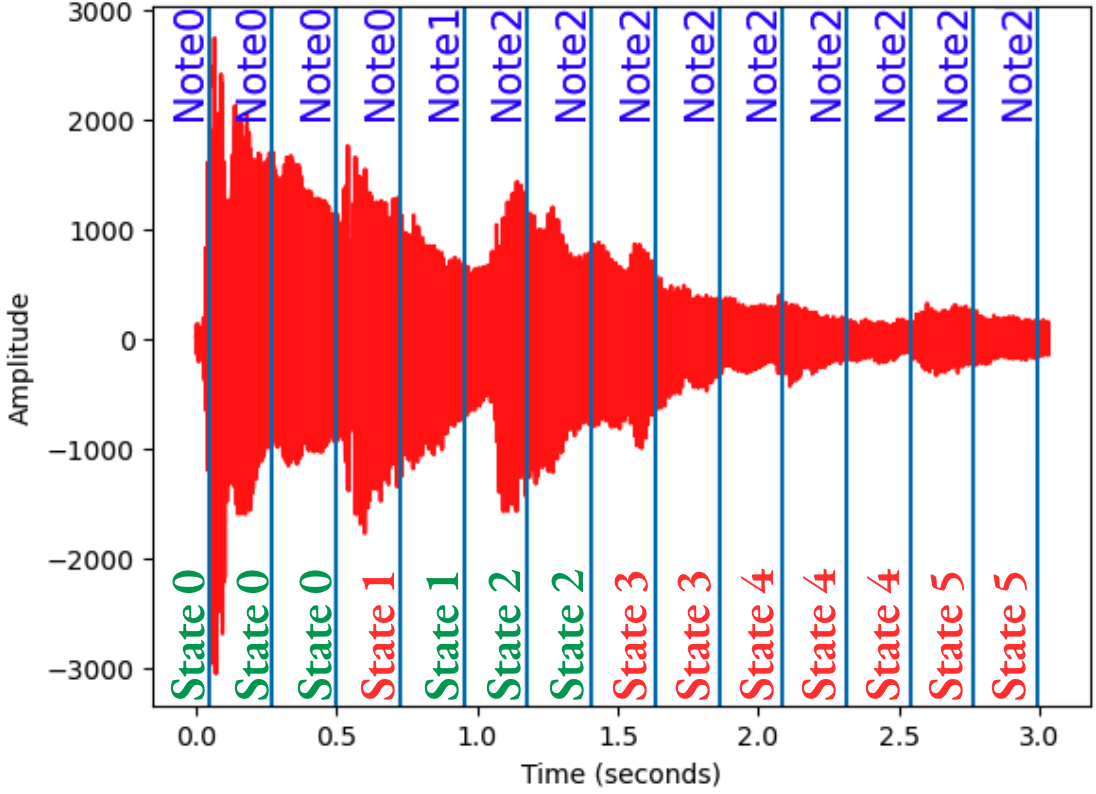}
    \caption{Excessive pedal and sharply decreasing loudness causing 8 mistakes.}
    \label{fig:loads_pedal}
  \end{subfigure}
  \caption{We show results from three different recordings of Bach's \textit{Prelude in C major BWV 846}. The blue lines represent the locations of each audioframe (where time between audioframes have been exaggerated to allow for visual inspection). The blue text represents predicted states by the Windowed Viterbi algorithm, and the red and green text below shows the true underlying states (green when correct, red when incorrect).}
  \label{fig:intermediate_results}
\end{figure}


Analysing the errors in \hyperref[fig:loads_pedal]{Figure \ref*{fig:loads_pedal}} revealed that each audioframe was misidentified as \textit{previous} note states. Given the excessive use of sustain pedal and rapid decrease in loudness, these errors are unsurprising, because pedal allows notes from previous note states to continue sounding, producing an audioframe that consists of notes from both previous and current note states. One way to mitigate this problem was through the observation that, since the frequencies from previous note states are not present in our hyperparameter $\boldsymbol{f}$, the sustain pedal was essentially adding extra noise. We therefore tried increasing the model noise, $\sigma_n$, which had some improvement on the overall results for pedalled piano pieces. However, considering that sustain pedal is \textit{intended} to cause pitch deviation from notes in the score, the problem should instead be addressed by refining our GP model so that it matches the pitches that actually sound, not just the notes in the score. \\

Hence, we attempted to account for pedal using data present in the MIDI file corresponding to the use of sustain pedal. In the original model, we only detect notes that are notated as ongoing, while in the new model, we additionally detect notes which are currently being sustained by the pedal. However, performers usually deviate from the exact pedal markings in a piece. We also tried adding notes from several consecutive preceding states to current ones to model a   configurable amount of pedal. However, both of these methods failed to provide consistent improvement, so we do not typically utilise them in our model (though there is the option to use the former, by setting a $\verb|sustain|$ parameter to $\verb|True|$).\\

In summary, it is clear that these results offer significant insight into the performance of our completed score follower.  If we constrain our recordings to moderate pedal, the initial aims of this section are met (\hyperref[section:aims_and_reqs_Alignment]{section \ref*{section:aims_and_reqs_Alignment}}), since we have a near-optimal sequence finder that         works well on simple pieces in typical circumstances. Further, the algorithm runs in real-time, at least for short recordings, showing promise and potential for a real-time score follower.     

\part{IV. Implementation}
\chapter{Implementation}{\label{ch:implementation}}
In this chapter, we present the implementation of the final product. We start by discussing how the four steps introduced in \hyperref[ch:high_level_approach]{chapter \ref*{ch:high_level_approach}} are integrated. We then outline the main system components of our score follower, presenting each as an independent, self-contained module. We then combine this into an overall system architecture and finally introduce the open-source score renderer used to display the score and evaluate the score follower.       


\section{Score Follower Framework Details}
Our score follower conforms to the high-level framework presented in \hyperref[section:score_follower_framework]{section \ref*{section:score_follower_framework}}. In step 1, two score features are extracted from a MIDI file (see \hyperref[subsection:midi]{subsection \ref*{subsection:midi}}), namely MIDI note numbers\footnote{\href{https://inspiredacoustics.com/en/MIDI_note_numbers_and_center_frequencies}{https://inspiredacoustics.com/en/MIDI\_note\_numbers\_and\_center\_frequencies}} (corresponding to pitch) and note onsets (corresponding to duration). In step 2, the audio is streamed (whether from a file or into a microphone) and audioframes that exceed some predefined energy threshold are extracted. Here, audioframes are groups of contiguous audio samples, whose length can be specified by the argument \verb|frame_length|, usually between 800 and 2000 samples. The period between consecutive audioframes can also be defined by the argument \verb|hop_length|, typically between 2000 and 5000 audio samples. In step 3, score following is performed via a `Windowed' Viterbi algorithm (see  \hyperref[subsection:adjusting_viterbi]{subsection \ref*{subsection:adjusting_viterbi}}) which uses the Gaussian Process (GP) log marginal likelihoods (LMLs) for emission probabilities (see \hyperref[section:state_duration_model]{section \ref*{section:state_duration_model}}) and a state duration model for transition probabilities (see \hyperref[section:state_duration_model]{section \ref*{section:state_duration_model}}). Finally, in step 4 we render our results using an adapted version of the open source user interface, \textit{Flippy Qualitative Testbench}.

\section{Following Modes}
Two modes are available to the user: Pre-recorded Mode and Live Mode. The former requires a pre-recorded $\verb|.wav|$ file, whereas the latter takes an input stream of audio via the device's microphone. Note that both modes are still forms of score following, as opposed to score alignment, since in each mode we receive audioframes at the sampling rate, not all at once.\\

Live Mode offers a practical example of a score follower, displaying a score and position marker which a musician can read off while playing. However, this mode is not suitable for evaluation because the input and results cannot be easily replicated. Even ignoring repeatability, Live Mode is not suitable for one-off testing since a musician using this application may be influenced by the movement of the marker. For instance, the performer may speed up if the score follower `gets ahead' or slow down if the position marker lags or `gets lost'. To avoid this, we use Pre-recorded Mode when evaluating the performance of our score follower. Furthermore, Pre-recorded Mode offers the advantage of testing away from the music room, providing the opportunity to evaluate a variety of recordings available online. 

\section{System Architecture}
Our guiding principle for development was to build modular code in order to create a streamlined system where each component performs a specific task independently. This structure facilitates easy testing and debugging. \hyperref[fig:black_box]{Figure \ref*{fig:black_box}} presents a high-level architecture diagram, where each black box abstracts a key component of the score follower. When operating in Pre-recorded Mode, there is the option to stream the recording during run-time, which outputs to the device's speakers (as indicated by the dashed lines).

\begin{figure}[H]
    \centering
    \includegraphics[width=1\textwidth]{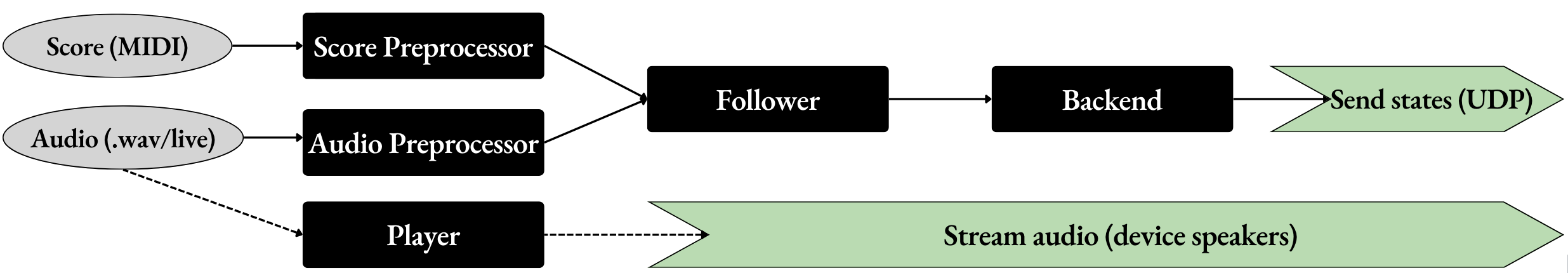}
    \caption{Abstracted system architecture diagram displaying inputs in grey, the 4 main components of the score follower in black and the outputs in green.}
    \label{fig:black_box}
\end{figure}

\subsection{Score Preprocessor}
The architecture for the Score Preprocesor is given in \hyperref[fig:score_preprocessor]{Figure \ref*{fig:score_preprocessor}}. First, MIDI note number and note onset times are extracted from each MIDI event. Simultaneous notes can be gathered into states and returned as a time-sorted list of lists called \verb|score|, where each element of the outer list is a list of simultaneous note onsets. Similarly, a list of note durations calculated as the time difference between consecutive states is returned as \verb|times_to_next|. Finally, all covariance matrices are precalculated and stored in a dictionary, where the key of the dictionary is determined by the notes present. This is because the distribution of notes and chords in a score is not random: notes tend to belong to a home \gls{key} and melodies tend to be repeated or related (similar to subject fields in speech processing). Therefore, states tend to be reused often, allowing us to achieve amortised time and space savings (by avoiding repeated calculation of the same covariance matrices). 

\begin{figure}[H]
    \centering
    \includegraphics[width=1\textwidth]{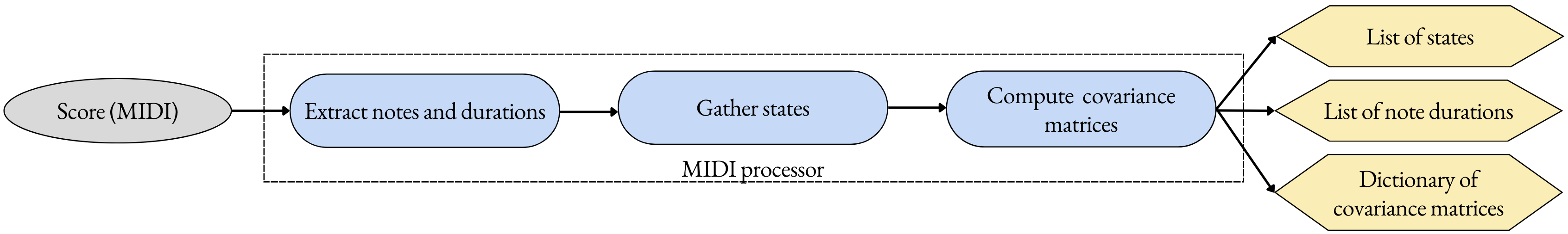}
    \caption{System architecture diagram representing the Score Preprocessor with inputs in grey, processes in blue and objects in yellow.}
    \label{fig:score_preprocessor}
\end{figure}

\subsection{Audio Preprocessor}
The architecture for the Audio Preprocessor is illustrated in \hyperref[fig:audio_preprocessor]{Figure \ref*{fig:audio_preprocessor}}. In Pre-recorded Mode, the Slicer receives a $\verb|.wav|$ file and returns audioframes separated by the \verb|hop_length|. These audioframes are periodically added to a multiprocessing queue, \verb|AudioFramesQueue|, to simulate real-time score following. In Live Mode, we use the python module \verb|sounddevice| to receive a stream of audio, using a periodic callback function to place audioframes on \verb|AudioFramesQueue|. 

\begin{figure}[H]
    \centering
    \includegraphics[width=1\textwidth]{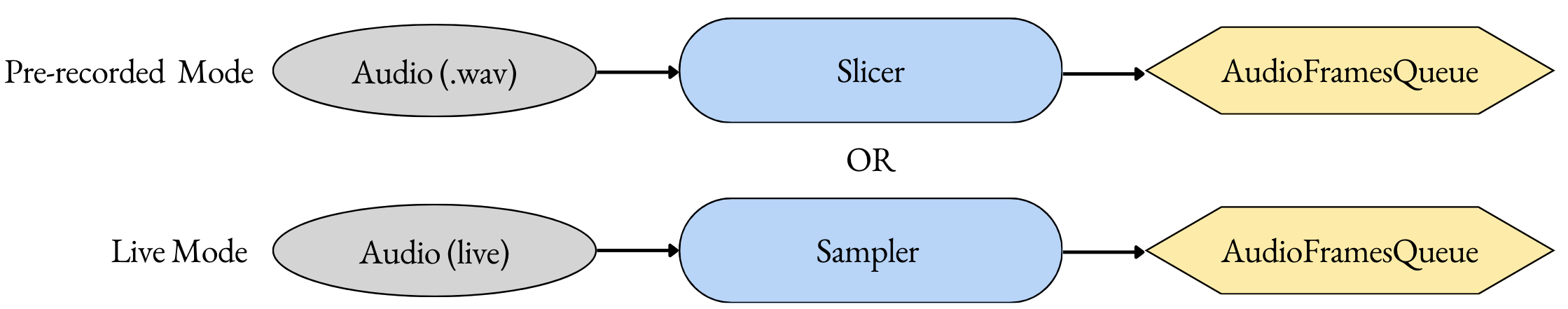}
    \caption{System architecture diagram representing the Audio Preprocessor with inputs in grey, processes in blue and objects in yellow.}
    \label{fig:audio_preprocessor}
\end{figure}

\subsection{Follower and Backend}
The joint Follower and Backend architecture diagram is shown in \hyperref[fig:follwer_and_backend]{Figure \ref*{fig:follwer_and_backend}}. The Viterbi Follower (detailed in \hyperref[subsection:adjusting_viterbi]{section \ref*{subsection:adjusting_viterbi}}) calculates the most probable state in the score, given audioframes continually taken from \verb|AudioFramesQueue|. These states are placed on another multiprocessing queue, the \verb|FollowerOutputQueue|, for the Backend to process and send. This prevents any bottle-necking occurring at the Follower stage. The Backend first sets up a UDP connection and then reads off values from \verb|FollowerOutputQueue|, sending them via UDP packets to the score renderer.

\begin{figure}[H]
    \centering
    \includegraphics[width=1\textwidth]{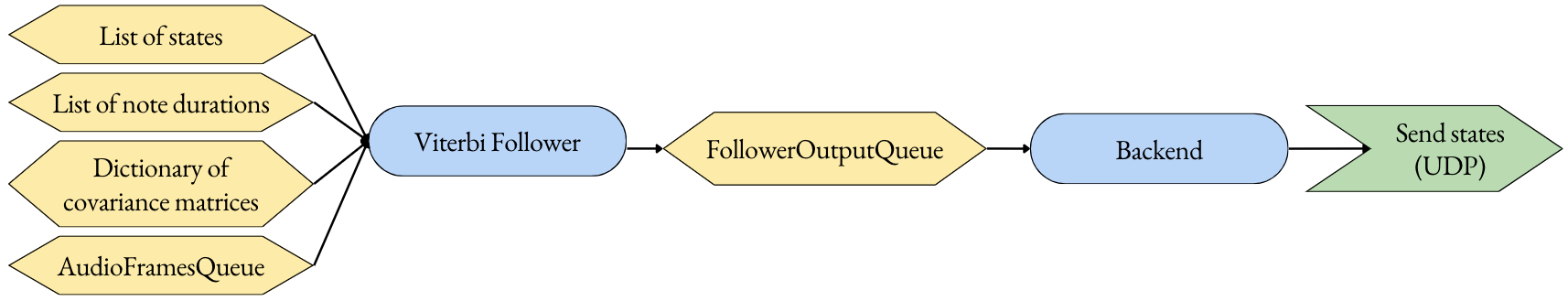}
    \caption{System architecture diagram representing the Follower and Backend processes with processes in blue, objects in yellow and outputs in green.}
    \label{fig:follwer_and_backend}
\end{figure}

\subsection{Player}
In Pre-recorded Mode, the Player sets up a new process and begins streaming the recording once the Follower process begins. This provides a baseline for testing purposes, as a trained musician can observe the score position marker and judge how well it matches the music. 

\subsection{Overall System Architecture}
The overall system architecture is presented in \hyperref[fig:overall_system_architecture]{Figure \ref*{fig:overall_system_architecture}}. Since the Follower runs a real-time, time sensitive process, parallelism is employed to reduce the total system latency. We use two \verb|multiprocessing| queues\footnote{\href{https://docs.python.org/3/library/multiprocessing.html}{https://docs.python.org/3/library/multiprocessing.html}} to avoid bottle-necking, which allows us to run 4 concurrent processes (Audio Preprocessor, Follower, Backend, and Audio Player). Hence, this architecture allows the components to run independently of one another to avoid blocking. Furthermore, this allows the system to take advantage of the multiple cores and high computational power offered by most modern machines.  

\begin{figure}[H]
    \centering
    \includegraphics[width=1\textwidth]{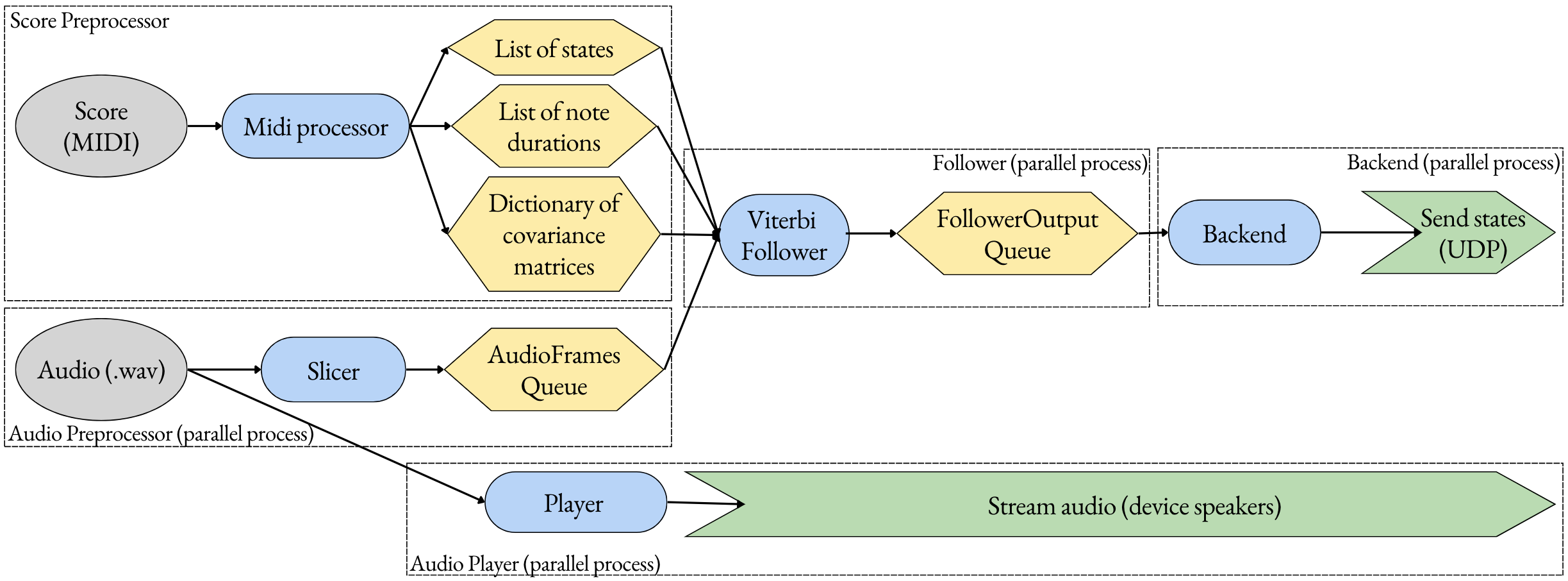}
    \caption{System architecture diagram representing the overall score follower running in Pre-recorded mode, with inputs in grey, processes in blue, objects in yellow and outputs in green.}
    \label{fig:overall_system_architecture}
\end{figure}

\section{Rendering Results}{\label{section:renderer}}
To visualise the results of our score follower, we adapted an open source tool for testing different score followers.\footnote{\href{https://github.com/flippy-fyp/flippy-qualitative-testbench/blob/main/README.md}{https://github.com/flippy-fyp/flippy-qualitative-testbench/blob/main/README.md}} \hyperref[fig:flippy_example]{Figure \ref*{fig:flippy_example}} shows the user interface of the score position renderer, where the green bar indicates score position. 

\begin{figure}[H]
    \centering
    \includegraphics{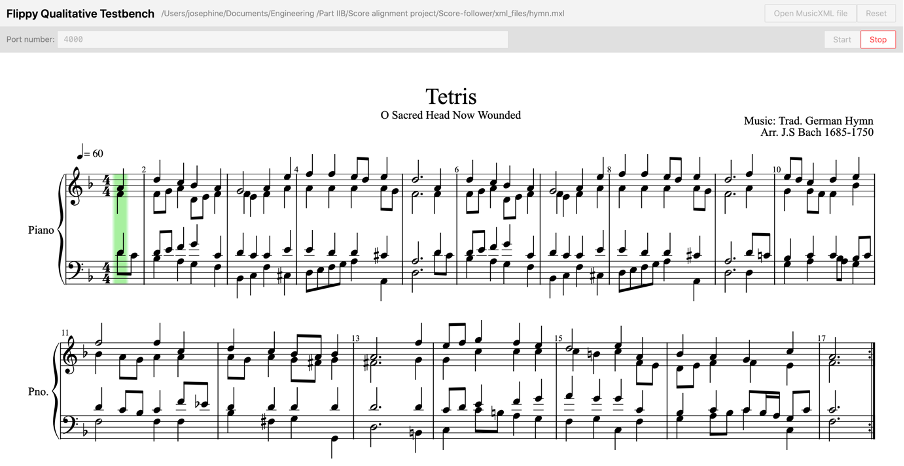}
    \caption{Screenshot of the score renderer user interface which displays a score (here we show a keyboard arrangement of \textit{O Haupt voll Blut und Wunden} by Bach). The green marker represents the score follower position.}
    \label{fig:flippy_example}
\end{figure}

\chapter{Results and Analysis}{\label{ch:results}}
To test the score follower, we collected a range of recordings of different pieces, varying in style, performer ability and instrument. We have largely avoided pieces with heavy use of \gls{sustain pedal} for the reasons explained in \hyperref[section:stage_2_results]{section \ref*{section:stage_2_results}}. Screen recordings of the first minute of each of these pieces being score-followed are hosted in the \verb|results| folder of the project repository.\footnote{\href{https://github.com/josephinecowley/Score-follower}{https://github.com/josephinecowley/Score-follower}} We have included the terminal window in our recordings, since this displays all input parameters at the beginning of each test, which is useful for repeating results. Furthermore, the terminal continually prints the log probabilities of notes so that results can be numerically analysed.

\subsubsection{\textit{Solfeggietto (H 220, Wq. 117: 2)} by C. P. E. Bach (anonymous performer)}
\textit{Solfeggietto} is a largely \gls{monophonic} piano piece. The performance of the follower was impressive, keeping up with the moderately fast, constant \gls{semiquaver} passages while never deviating more than one semiquaver from the true score location. Despite the performer in the recording not sustaining the left-hand notes in \gls{bars} 16 for their full duration, the score follower was nonetheless able to follow accurately using the other notes.  

\subsubsection{\textit{Minuet in G major, BWV Anh. 114} by J. S. Bach (anonymous performer)}
\textit{Minuet in G} is a simple two-part piano piece. Even after the addition of a second part, the follower succeeded in the first half of the piece. Despite getting lost due to the omission of a left hand note, the score follower was ultimately able to recover by the end of the piece. Furthermore, it was never more than one bar behind, which is roughly sufficient for many applications (including automatic page turning— refer to \hyperref[subsection:APT]{subsection \ref*{subsection:APT}}).  It was necessary to set the \verb|sustain| parameter to \verb|True| since there were locations (such as bar 2) where notes in the left hand were sustained whilst the right hand changed. Without $\verb|sustain|=\verb|True|$, the score follower fails. 

\subsubsection{\textit{O Haupt voll Blut und Wunden} from \textit{St Matthew Passion Mvt. 54 BWV 244} by J. S. Bach, arranged for keyboard (played by author)}
\textit{O Haupt voll Blut und Wunden} is a 4-part German traditional hymn. We have used a Bach keyboard arrangement which preserves its 4-part \gls{homophonic} \gls{texture}. Despite this texture and a moderate amount of pedal, the score follower succeeded without significant error.

\subsubsection{\textit{Syrinx, L. 129} by Claude Debussy (played by Emmanuel Pahud)}
\textit{Syrinx} is a solo flute piece. Although the GP model was designed for the piano, we use this recording to test its performance on a range of instruments. The following of this piece was highly accurate, showing the flexibility of the GP model. For this piece, which has high \gls{rhythm}ic freedom and therefore lots of rubato and many fermatas (i.e. pauses), we disabled the \verb|state_duration| mode, as the false assumption of locally constant tempo caused the follower to `get lost'. We have included this second test as an example of the limitations of our state duration model. In that recording the score follower gets especially lost in bar 4 due to the rapidly fluctuating tempo.

\subsubsection{\textit{Allemande} from \textit{Partita in A minor for solo flute, BWV 1013} by J. S. Bach, arranged for oboe (played by Bernice Lee)}
\textit{Allemande, BWV 1013,} is a solo flute work, though here we use a performance on the oboe for a wider range of test cases. The follower primarily achieves its goal, demonstrating the remarkable ability of the follower to keep to performances which exhibits much \gls{rubato}. In the final bar of the recording, however, the follower becomes `lost' due to the added \gls{ornament}. This demonstrates a limitation of such strict followers.  

\subsubsection{\textit{Allemande} from \textit{Partita in D minor No. 2 for Violin, Allemande, BWV 1004} by J. S. Bach (played by Itzhak Perlman)}
\textit{Allemande, BWV 1004,} is a solo violin piece. The program accurately follows the piece, impressively keeping up with Perlman's heavy use of rubato and fast runs of \gls{demisemiquavers}. Despite the fact that the violin produces pitches which are not perfectly in tune, both intentionally due to vibrato and unintentionally due to mistakes (e.g., the highest note in bar 14, which is flat), the score follower was able to match almost all notes accurately.


\subsubsection{\textit{Adagio and Allegro, Op. 70} by Robert Schumann (played by Christina Bensi)}{\label{subsubsection:Christina}
\textit{Adagio and Allegro} is a duet for piano and horn in F, but we use a performance of a cello-piano duet. As a multi-instrument duet, with a thick piano accompaniment which utilises generous sustain pedal, this piece truly exposes the limits of our current score follower. The follower rarely positions the marker on exactly the right state/note, though in general, it does succeed in `hanging on', at least for the first 12 bars. This demonstrates that even though the score follower may not be positioned on the correct state, it can recover if using a wide enough window in the Windowed Viterbi algorithm. Note that we turned off \verb|state_duration| for this example to show it was truly the GP model and not just the state duration model causing the follower to proceed at the average pace of the underlying tempo. Additionally, we found increasing the model noise, $\sigma_n$, improved the accuracy, since sustain pedal effectively adds noise (as explained in \hyperref[section:stage_2_results]{section \ref*{section:stage_2_results}}).  \\

Besides the Viterbi algorithm allowing recovery to the correct state within a certain window (see \hyperref[subsection:adjusting_viterbi]{section \ref*{subsection:adjusting_viterbi}}), part of the reason the follower could `hang on' was because the underlying \gls{harmonic progression} of the piece changes slower than the individual note states. These notes states are statistically linked to these `higher level states' (i.e. underlying chords and \gls{key}s) which means that the score follower is usually more likely to go to the correct \textit{area}, even if not precisely the right state. Put in terms of music theory, the score follower is able to follow harmonic changes, even when following exact notes is not possible, which mirrors the human score-following techniques explored in \hyperref[subsection:manual_score_following]{subsection \ref*{subsection:manual_score_following}}. An interesting area for future work could be the use of multi-resolution HMMs to capture the hierarchical structure of music across different time scales \cite{Baggenstoss}.

\section{Key Findings and Limitations}
 These results show that we have built a largely successful score follower, which is capable of following intermediate monophonic, homophonic and 4-part harmony piano pieces. Additionally, despite being modelled for piano, the follower is arguably better at following instruments including the flute, oboe and violin, yielding excellent results. This not only reflects the simpler task of following these one- or few-line instruments with more uniform temporal envelopes, but also the versatility of GPs. On all instruments, the follower adapted impressively well to heavy rubato and changes in tempo.   \\
 
One difficulty is piano with heavy sustain pedal, since it is impossible to predict note sustain lengths. Additionally, the model fails at distinguishing loud note sources from quiet ones (as explored in \hyperref[subsubsection:amplitude]{subsubsection \ref*{subsubsection:amplitude}}), which means that the relatively quiet notes that continue to sound due to sustain pedal continue to be detected. Another limitation is performer embellishment and other deviations from the score. Finally, passages of repeated notes or phrases shorter than the Viterbi window sometimes present issues, since the follower cannot distinguish between instances of repetitions, in better cases exhibiting small jumps, and at worst becoming `lost'. Outside of these challenges, however, the score follower performs impressively.

\part{V. Discussion }
\chapter{Summary and Conclusion}{\label{ch:conclusion}}

\section{Summary}
In Parts I and II of this paper, we defined the problem of score following and outlined the methodology we would use to develop our novel approach to score following using GPs.
Part III gave the statistical models underlying our solution, and in Part IV, we described our implementation and showed that the solution performed highly promisingly. Overall, we have achieved successful score following using Gaussian Processes (GPs), therefore fulfilling the original project goal set out in \hyperref[section:project_goal]{section \ref*{section:project_goal}}. Since the novel work of this paper occurs in chapters 7 - 9, we proceed with a brief summary of each.\\

In \hyperref[ch:model_selection]{chapter \ref*{ch:model_selection}} (Stage 1: \textit{GP Model Specification}), we developed a GP model for the piano, which was both accurate and flexible. This GP model is novel in its use of a Spectral Mixture kernel, which  allows encoding the spectral envelope of pitched instruments. We used empirical optimisation techniques to determine values for the covariance function's hyperparameters. We then used the GP model to build an efficient and stable log marginal likelihood (LML) function of various notes given an audioframe.\\


In  \hyperref[ch:alignment]{chapter \ref*{ch:alignment}} (Stage 2: \textit{Real-Time Alignment}), we designed a fairly robust algorithm for online score following. This included a `Windowed' Viterbi algorithm, using the LMLs from Stage 1 for the emission probabilities, and a state duration model for the transition probabilities.  \\

Finally, in \hyperref[ch:implementation]{chapter \ref*{ch:implementation}}, we built a powerful product which uses a score and recording (or live performance) to perform score following. This product proved especially effective for un-pedalled piano pieces, for a wide range of textures, and readily adapted to changes in \gls{tempo} and exaggerated \gls{rubato}.  Indeed, in some ways we have exceeded our project goal, since the product could follow other solo instruments, and even displayed the beginnings of score following for complex multi-instrument pieces. The smooth running of the product, which demonstrated low latency and no buffering, proved significant in light of initial scepticism regarding the computational complexity of GPs. This speaks for the careful choice of system architecture and effective use of multiprocessing.\\

Hence, we have successfully provided a proof of concept that GPs can be harnessed as an effective method for score following. Note that by definition this also means that GPs are capable of score alignment, the off-line counterpart of score following (refer to \hyperref[subsection:score_following_v_alignment]{section \ref*{subsection:score_following_v_alignment}}).

\section{Areas for Future Research}

\subsubsection{Other Instruments and Ensembles}

One simple extension to this project would be to develop specific GP models for other instruments. Given the impressive results of our GP model (which was designed for piano) on other instruments, we conjecture that it would be feasible to create such models by adjusting the overtones as well as accounting for new physical effects (e.g., perhaps inharmonicity effects are different on other types of instruments).\\

Then, we could selectively use the different models to simultaneously follow separate parts in pieces with multiple instruments, making each follower’s job simpler, building in robustness, and shifting the focus from stationary single audioframes to melody lines, moving towards a more `human way' of score following (as outlined in \hyperref[section:score_following_approaches]{section \ref*{section:score_following_approaches}}). This approach would therefore make use of \gls{timbre} information, as well as pitch and duration. Of course, methods of source separation would need to be investigated.

\subsubsection{Improving Polyphonic Score Following}
The idea in the previous paragraph could be used to tackle the challenge of \gls{polyphonic}, where there are distinct melodies simultaneously playing (e.g., a fugue).   

\subsubsection{Multi-Resolution Hidden Markov Models (MRHMMs)}
As suggested in \hyperref[ch:results]{chapter \ref*{ch:results}}, MRHMMs could be used to encode the hierchical structure of music. As well as musical notes as a low level latent state, \gls{harmonic progression}s could serve as a coarser resolution for the MRHMM \cite{Baggenstoss}. MRHMM segments could be obtained from the score (e.g., guitar tabs) or could be calculated at run time.

\subsubsection{Adaptive Training}
To avoid the extensive pre-processing and training of the models, one interesting area of research could be incorporating adaptive models which optimise hyperparameters in real-time. 

\subsubsection{Related Applications}
The success of this investigation shows the potential that GPs offer in the wider context of Music Information Retrieval (MIR) tasks and more general practical engineering applications. Some ideas include using GPs for tuning and pitch detection, or instrument identification.


\section{Final Reflections and Conclusion}
Automatic score following is no easy task, as demonstrated by the lack of uptake among musicians despite substantial research over the last four decades. By contrast, this project represents only a foray into score following over just eight months. However, our results provide a meaningful contribution to the field by showing that GPs are a viable method for statistical inference that fits into existing Bayesian frameworks for score following. Our successful results have highlighted the power and flexibility of GPs and how they can be used in practical engineering applications. Future work, with particular attention to the few key sources of error as well as user experience, could help better utilise the algorithms developed in this project to produce an application suitable for use by amateur and professional musicians alike, from the practice room to the concert hall. 

\appendix
\begin{appendices}
\chapter{}
\vspace{-1.8cm}
\setstretch{0.869}
\section{Risk Assessment Retrospective}

Given that the project was entirely computational, risks outlined in the initial hazard assessment were primarily computer based, covering hazards caused by excessive or improper use of computer based equipment, namely eye strain and repetitive stress on the body caused by bad posture and extended periods of writing or coding. In retrospect, these issues were largely avoided by the investment of a monitor at the beginning of the year, and ensuring suitable supporting computer based equipment.

\section{Log Book}

Throughout this project, we used \href{https://toggl.com/}{Toggl} to record details of progress and hours spent working. Our Toggl report can be accessed on \href{https://drive.google.com/file/d/1Inj8odZ7XiuMx6RSc6DPxtUZ8Bg0fgdk/view?usp=sharing}{Google Drive}. 

\section{Final Product}

Source code for the score follower developed in this project is available on \href{https://github.com/josephinecowley/Score-follower}{GitHub}. It is designed for use with the adapted score renderer app available \href{https://github.com/josephinecowley/Score-follower/releases/tag/Release}{here}. 

\section{Inverse Fourier Transform of Covariance Function}{\label{appendix:iFT}}

\footnotesize
First, define the Fourier transform as $G(f) = \int_{-\infty}^\infty g(t) e^{-j f \tau} \, d\tau $. Note that we  use this form of the Fourier transform to be consistent with \cite{wilson_2013_gaussian}.
The inverse Fourier transform (iFT) of the Gaussian $\phi(f; 0, \sigma)$ is:
\begin{align*} 
k(\tau) 
&= \medint\int_{-\infty}^{\infty} \phi(f)e^{j 2\pi f\tau} \ df \\
&= \mfrac1 {\sigma \sqrt{2\pi}}\medint\int_{-\infty}^{\infty}e^{-\frac{f^2}{2\sigma^2} + j2\pi f\tau} \ df \\
&= \mfrac1 {\sigma \sqrt{2\pi}}\medint\int_{-\infty}^{\infty}e^{-\frac{1}{2\sigma^2} \big[(f-j2\pi \sigma^2 \tau)^2 + 4\pi^2 \sigma^4 \tau^2 \big]} \ df \\
&= \mfrac1 {\sigma \sqrt{2\pi}}\medint\int_{-\infty}^{\infty}e^{-\frac{1}{2\sigma^2}(f-2\pi \sigma^2 \tau)^2 - 2\pi^2 \sigma^2 \tau^2} \ df 
\\ & \ \ \ \  \text{Let } p = \mfrac{f-2\pi \sigma^2 \tau}{\sqrt 2 \sigma}. \text{ Then } df = \sqrt2 \sigma \  dp. \\
&= \mfrac {e^{- 2\pi^2 \sigma^2 \tau^2}} {\sigma \sqrt{2\pi}}\medint\int_{-\infty}^{\infty} e^{-p^2} \sqrt 2 \sigma \ dp \\
&= \mfrac {e^{- 2\pi^2 \sigma^2 \tau^2}} {\sqrt{\pi}}\medint\int_{-\infty}^{\infty} e^{-p^2} \ dp \\
&= e^{-2\pi^2 \sigma^2 \tau^2} 
\end{align*}
Hence, using the frequency shift property of Fourier transforms: $iFT(\phi(f; m f_q, \sigma^2_f )) =  e^{-2\pi^2 \sigma_f^2\tau^2} e^{2\pi j m f_q \tau}$. Observing the linearity of the Fourier transform, the fact that $cos(x) = \frac{e^{jx} + e^{-jx}}{2}$, and multiplying $f_q$ by the inharmonicity constants $b_{m,f_q}$, we show that the covariance function of \hyperref[equation:frequency]{equation \ref*{equation:frequency}} is: 
\begin{align*}
k(\tau) = e^{-2\pi^2\sigma_f^2 \tau^2} \sum_{q=1}^Q w_q \sum_{m=1}^M E_m \cos(2\pi m f_{q} b_{m,f_q}  \tau) \qed
\end{align*}
\end{appendices}

{\footnotesize
\setstretch{0.7}
\printbibliography[heading=bibintoc]
}

\end{document}